\documentclass[11pt,titlepage]{article}

\usepackage{graphicx}
\usepackage{endfloat}
\usepackage{amsmath,amsfonts}
\usepackage{mychicago}
\usepackage{subfigure}
\usepackage{genetics_manu_style}
\usepackage{color}
\usepackage{soul}
\usepackage{lipsum,appendix}

\graphicspath{{./Figures/}}

\title{Stability and response of polygenic traits to stabilizing selection and mutation}

\author{Harold P. de Vladar\thanks{hpvladar@ist.ac.at},  Nick Barton\thanks{Nick.Barton@ist.ac.at}}

\renewcommand{\CorrespondingAddress}{IST Austria \\
    Am Campus 1 \\ 
    Klosterneuburg 3400\\ 
    Austria}
\renewcommand{\RunningHead}{Stabilizing selection on polygenic traits}
\renewcommand{\CorrespondingAuthor}{H.P. de Vladar}
\renewcommand{\KeyWords}{genetic variance; distribution of allelic effects; moving optimum.}

\newcommand{\citet}{\citeN}
\newcommand{\citep}{\cite}

\begin{document}

\maketitle

\begin{abstract}When polygenic traits are under stabilizing selection, many different combinations of alleles allow close adaptation to the optimum.
If alleles have equal effects, all combinations that result in the same deviation from the optimum are equivalent. Furthermore, the genetic variance
that is maintained by mutation-selection balance is $2 \mu/S$ per locus, where $\mu$ is the mutation rate and $S$ the strength of stabilizing selection.  In reality, alleles vary in their effects, making the fitness landscape asymmetric, and complicating analysis of the equilibria. We show that that the resulting genetic variance depends on the fraction of alleles near fixation, which contribute by $2 \mu/S$, and on the total mutational effects of alleles that are at intermediate frequency. The interplay between stabilizing selection and mutation  leads to a sharp transition: alleles with effects smaller than a threshold value of $2\sqrt{\mu / S}$  remain polymorphic, whereas those with larger effects are fixed. The genetic load in equilibrium is less than for traits of equal effects, and the fitness equilibria are more similar. We find that if the optimum is displaced, alleles with effects close to the threshold value sweep first, and their rate of increase is bounded by $\sqrt{\mu S}$. Long term response leads in general to well-adapted traits, unlike the case of equal effects that often end up at a sub-optimal fitness peak. However, the particular peaks to which the populations converge are extremely sensitive to the initial states, and to the speed of the shift of the optimum trait value.
\end{abstract}
\section*{Introduction}

Understanding quantitative genetics in terms of population genetics is crucial for both scientific and practical reasons. However, the development of a consistent theory for long term evolution has had limited success, because the polygenic basis of quantitative traits makes the prediction of their response to selection immensely intricate, even under the simplest assumptions (e.g. additivity, equal effects of an allele on the trait, and linkage equilibrium) \citep{Barton:1989du,Keightley:1990vb,Turelli:1994ta}. Most traits seem to be under some form of stabilizing selection, either by the direct action of selection on a trait whose extreme values are unfit, or indirectly by compromising individual fitness due to pleiotropic detrimental effects \citep{Keightley:1990vb,Mackay:2001ee,Hill:2012ks,Mackay:2010je}.
The joint effects of stabilizing selection and mutation lead to very complicated  allele frequency equilibria and evolution, and it is not
obvious how much genetic variation they can maintain \citep{Turelli:1984tb,Turelli:1988tm,Burger:2001wy}.

An exact analysis in terms of allele frequencies is lacking for polygenic traits with loci of unequal effects. This is desirable, as data from genome wide association studies (GWAS) yield  information about the distribution of single nucleotide polymorphisms (SNPs) relevant to several traits based on  population sequence data  \citep{Hindorff:2009cc,Visscher:2012je}. This makes it urgent to understand how variation at the molecular level explains phenotypic variation. How quantitative variation depends on the number of loci and on the distribution of allelic effects is not  clear, and is the central question of this article.

\begin{figure}[t]
\begin{center}
\includegraphics[scale=0.32]{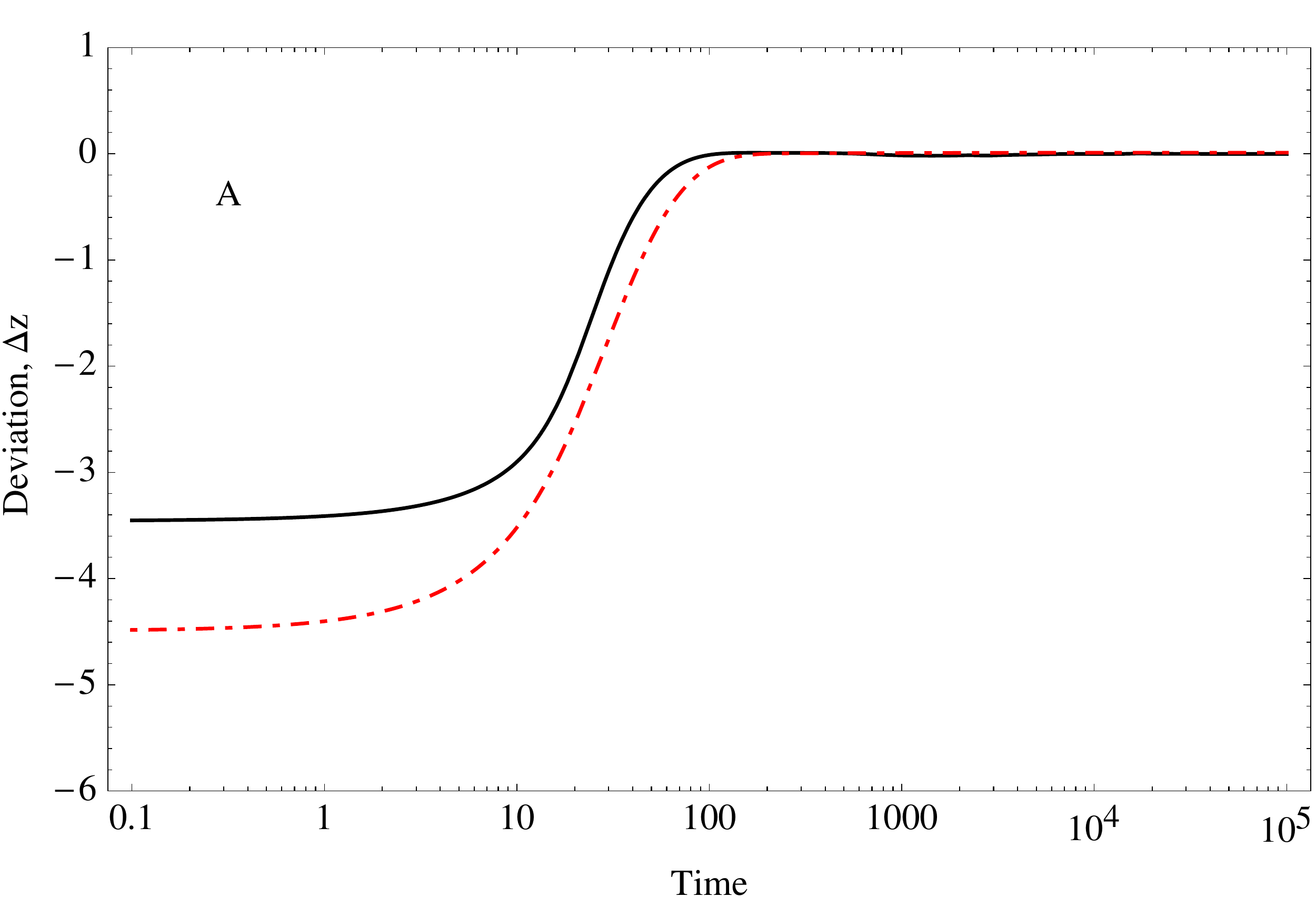}
\includegraphics[scale=0.32]{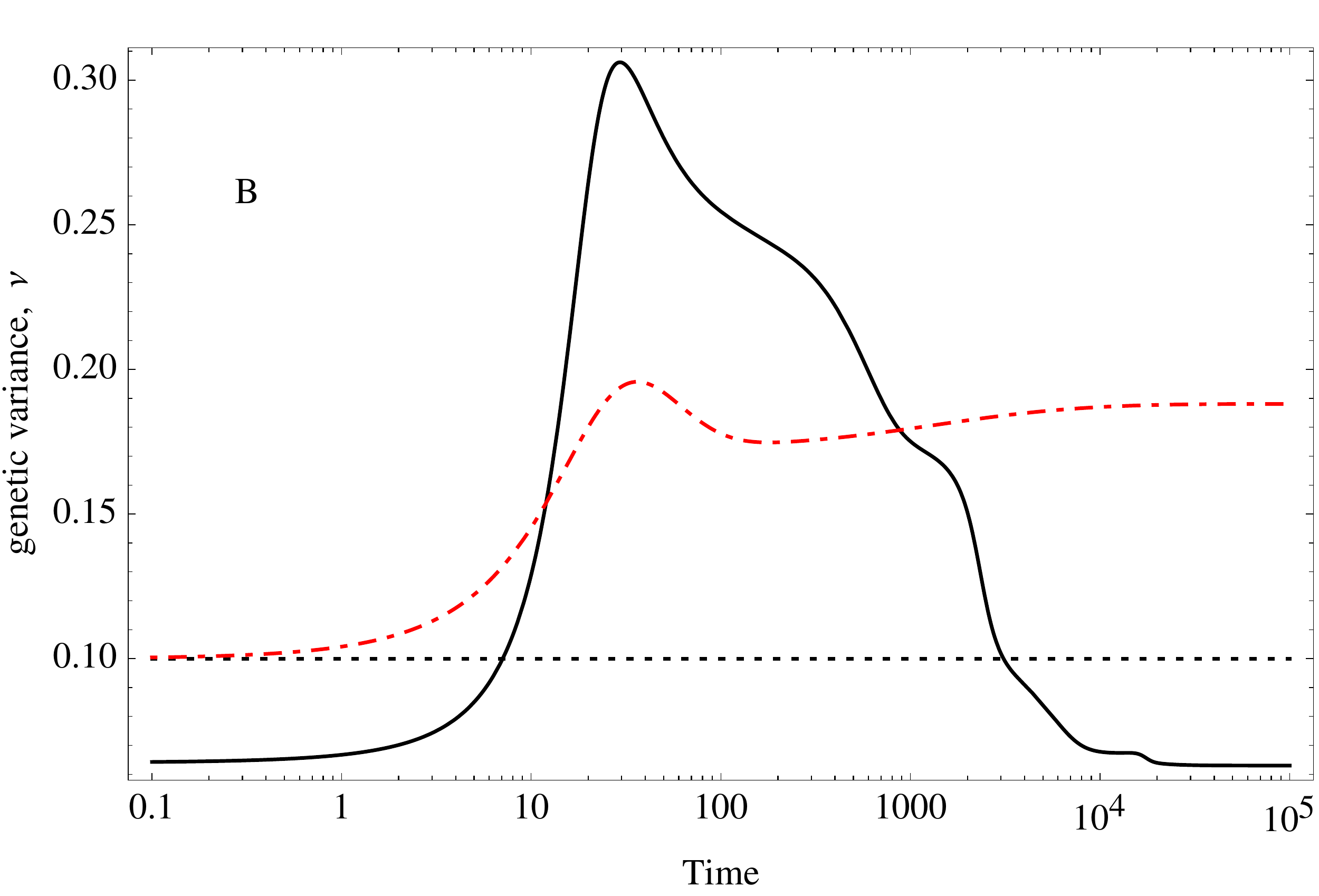}
\includegraphics[scale=0.32]{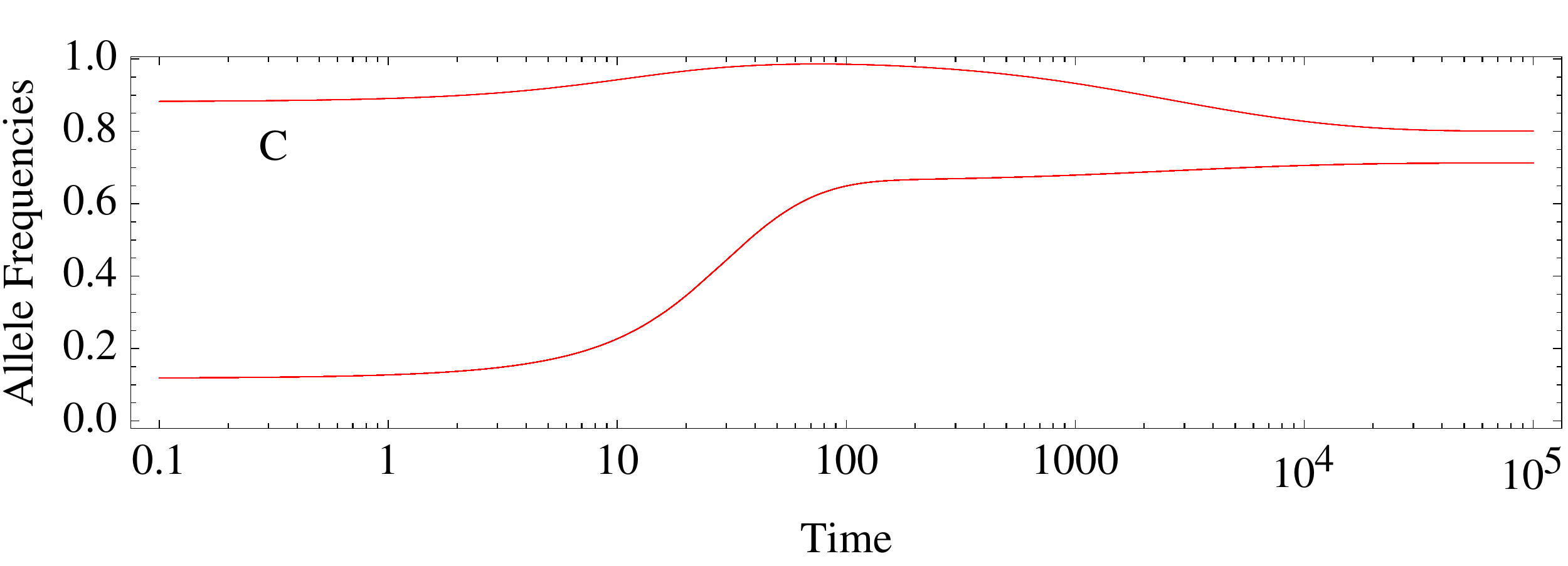}
\includegraphics[scale=0.32]{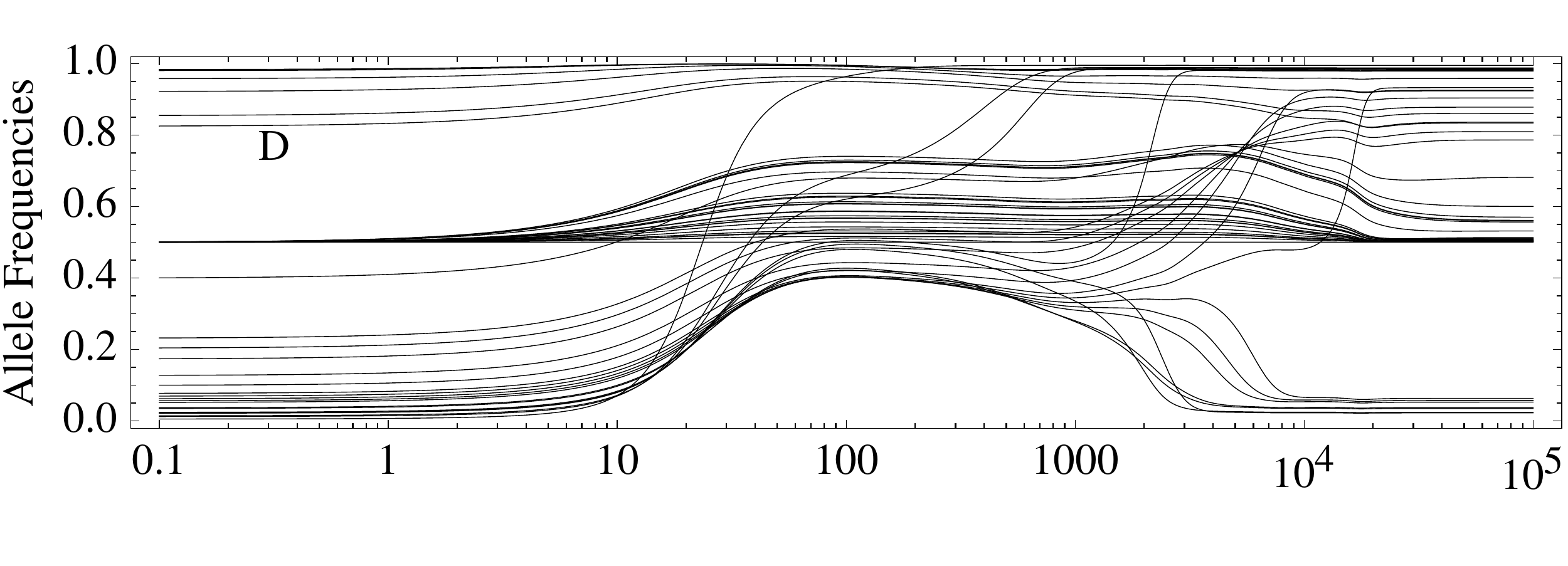}
\end{center}
\caption{Response to selection of traits determined by 50 loci of equal (red and dash-dotted, color online) and unequal effects (black).
(A) Deviation of the trait mean from the optimum value. (B) Genetic variance.  The dotted black line show the HoC variance.  
Allele frequencies under (C) equal and (D) unequal effects. The equal effects have $\gamma =1/10$, and the unequal effects are distributed as an exponential (mean=1/10).  Mutation rate $\mu =10^{-4}$, selection intensity $S=10^{-1}$. The dynamics are numerical solutions to ordinary differential equations (Eq.  6  in the text).
}
\label{Fig:EqvsUneqDyn}
\end{figure}

\citet{Barton:1986vl} showed that when a population is at equilibrium with the trait mean is at the optimum and the alleles are very close to fixation, then the genetic variance, $\nu$, is $2n\mu/S$, where $n$ is the number of contributing loci, $\mu$ is the per locus mutation rate and $S$ is the strength of stabilizing selection. In general, the genetic load $L$ is due to both deviation from the optimum and genetic variance, $L\propto \Delta z^2 + \nu$, where $\Delta z$ is the deviation of the trait mean from the optimum. The contribution of the genetic variance is much more significant because compared to it, the deviations from the optimum are very small. Moreover, if the trait is not at the optimum, higher variance can be maintained. These calculations assumed a trait with diallelic loci of equal effects. However, we will show that under unequal effects, deviations from the optimum can also maintain less variance.

The response to a shift in the optimum trait value will be radically different under equal and unequal effects. For example, Fig. \ref{Fig:EqvsUneqDyn} shows the response of two equivalent populations that differ only in their distribution of allelic effects. Notice that although the traits match the optimum almost perfectly in both cases (Fig. \ref{Fig:EqvsUneqDyn}A), under equal effects much more variation is maintained than under unequal effects (Fig. \ref{Fig:EqvsUneqDyn}B), which implies a greater mutation load.

We will see that under unequal effects, the equilibria depend on the magnitude of allelic effects. With equal effects, there is a high degree of symmetry in the sense that many allelic combinations match a given optimum value, making it easier to characterize the possible equilibria \cite{Barton:1986vl}. However, this analysis fails under unequal effects because the symmetry is absent. For example, Figs. \ref{Fig:EqvsUneqDyn}C,D show the response of the allele frequencies; in C and D the alleles have equal and unequal effects, respectively. Initially, the alleles rest at a stable equilibrium that have comparable mean and variance. We see that the response under unequal effects is more heterogeneous in Fig. \ref{Fig:EqvsUneqDyn}D (unequal effects), whereas the alleles respond homogeneously under equal effects (Fig. \ref{Fig:EqvsUneqDyn}C). This difference in the response accounts for the eventual mal-adaptation of traits with loci of equal effects. Moreover, alleles that have very large effects are at very low frequency and might take substantial time to achieve a higher representation in the population. Thus, anticipating when they will  reach intermediate frequencies and make a notable contribution to the genetic variance is difficult. Also, we don't know which alleles contribute preferentially to the response and eventual adaptation of the trait. Under equal effects, all alleles have the same contribution, but the symmetry of the solutions effectively reduces the genetic degrees of freedom, which in turn limits the possible paths to find a global fitness optimum.

The House of Cards  (HoC) is a mutation-selection balance model that assumes that each allele is new and arises independently from the previous allele from which it mutated, so that the effects of new mutations are uncorrelated from the previous ones. In equilibrium, the variance of allelic effects is larger than the genetic variance, and is predicted to be $2n\mu/S$, where $n$ is the number of loci, $\mu$ is the per-locus mutation rate, and $S$ is the strength of stabilizing selection \citep[, Ch. IV]{Turelli:1984tb,Kingman:1978tq,Burger:2001wy}. Exactly this amount of genetic variance is maintained for traits with several diallelic loci of equal effects, when they are adapted to the optimum  \citep{Barton:1986vl}. However, numerical experiments such as the one shown in Fig. \ref{Fig:EqvsUneqDyn}B  reveals that with unequal effects, the genetic variance is decreased even further below this bound. It is not immediately clear why this difference between traits with equal and unequal effects occurs.

\begin{figure}[t]
\begin{center}
\includegraphics[scale=1.1]{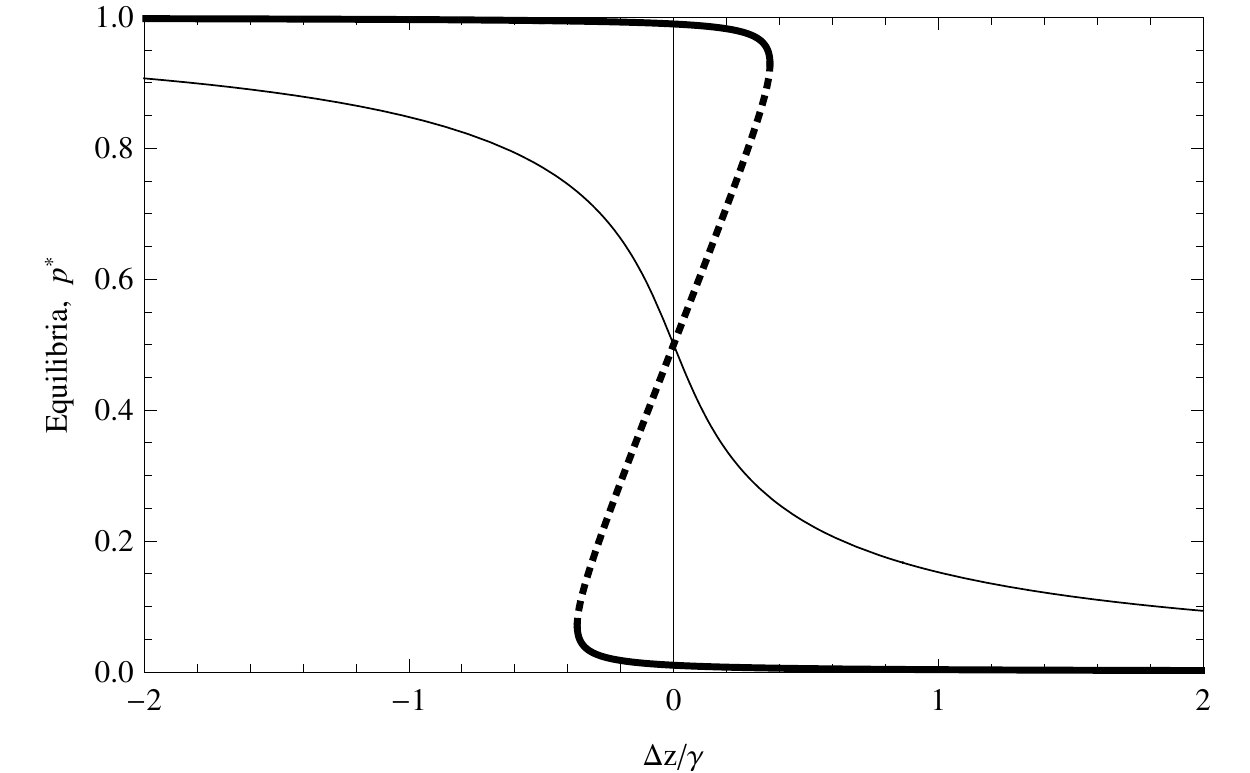}
\includegraphics[scale=1.1]{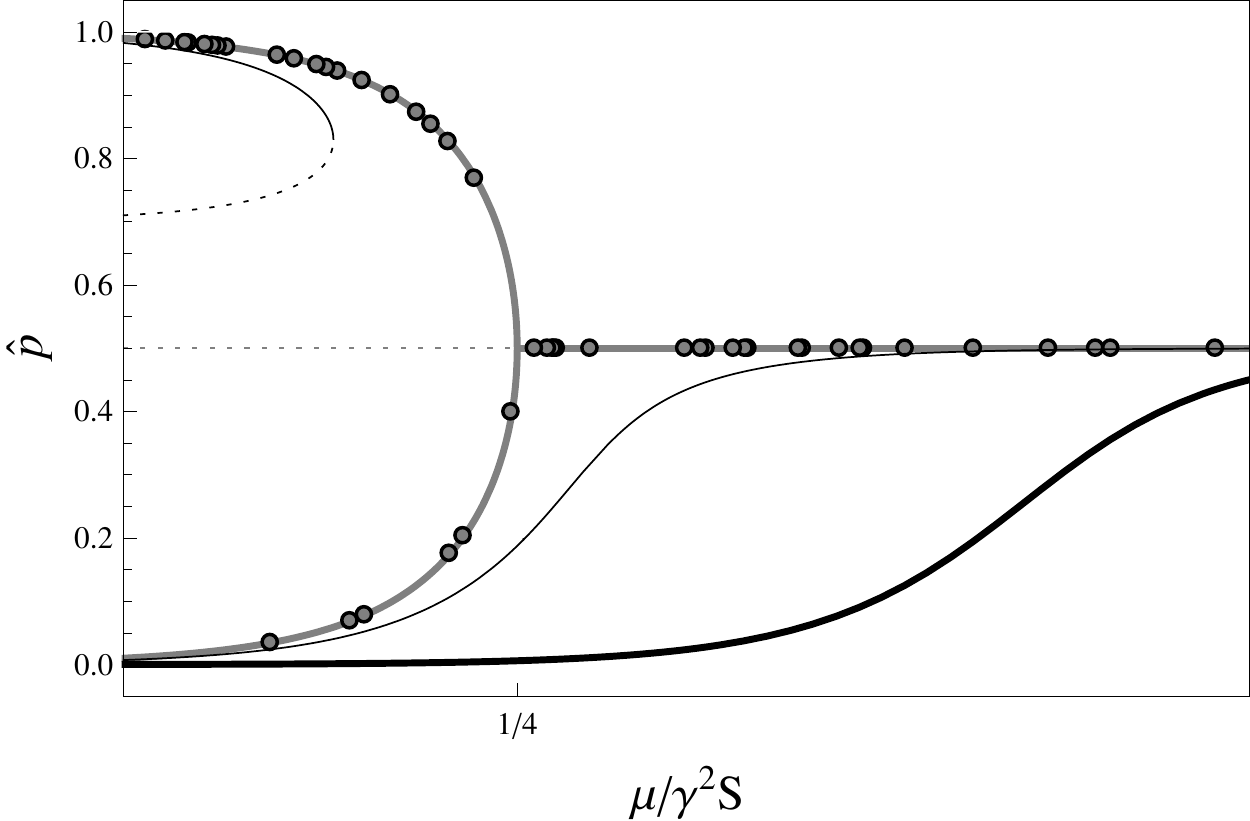}
\end{center}
\caption{ (A) Equilibria of allele frequencies as a function of  scaled deviation from the optimum. Thin curve: equilibria for alleles of small effects $\left(\gamma ^2<4\mu /S \right)$;   for each value of $ \Delta z/\gamma $ there is one stable allele frequency. Thick curve: equilibria for alleles of large effects, $ \left(\gamma ^2>4\mu /S \right) $;  for small deviations from the optimum, there are two possible equilibria near fixation. the dashed segment are unstable equilibria.
(B) Equilibria of allele frequencies as a function of the scaled parameter $ m=\mu \left/\gamma ^2\right.S $.  Thick gray curve: no deviations from the optimum, $\Delta z$=0. Thin black curve: small deviations from the optimum. Black thick curve:  large deviations from the optimum.
Circles: end point of Fig. \ref{Fig:EqvsUneqDyn}D. }
\label{Fig:BifurcationZM}
\end{figure}

Although the HoC assumes a continuous production of new alleles with varying effects \cite{Turelli:1984tb}, this model can be interpreted as a limit of a trait constituted of many loci  \citep{Barton:1986vl}, in which case each locus composing a trait $z$ under stabilizing selection evolves according (as will be explained in detail below) to:
\begin{equation} 
\frac{dp}{dt}=-S \gamma  p(1-p)(2\Delta z+\gamma  (1-2p))+\mu (1-2p) ~,
\end{equation}
where $p$ is the frequency of the `+' allele, $\gamma $ is its allelic effect, and $\Delta z$ the deviation of the trait mean from the optimum. Detailed analyses of this system under equal effects were performed by \citet{Barton:1986vl}.  At equilibrium, the equation above can have one or three solutions for each locus, given by the cubic polynomial on
$p$ that results from equating \(\frac{dp}{dt}=0\). If we plot the equilibrium value of $p$ against
$\Delta z$/$\gamma $ we find that there are two types of curves, depending on the mutation rate, $\mu $ (Fig. \ref{Fig:BifurcationZM}A). The first type (Fig. \ref{Fig:BifurcationZM}A, thin
curve) occurs at high mutation rates; in this case and at small deviations from the trait optimum, the equilibrium is maintained at intermediate
frequencies, maintaining substantial variability. The other type of equilibrium (Fig. \ref{Fig:BifurcationZM}A, thick curve) occurs when mutation rates are low compared
to the mutational effects: for well-adapted traits either of the two alleles can be near fixation, each one contributing to the genetic variance
by 2$\mu $/$S $,  as the HoC predicts.

Under equal effects the equilibrium value of the trait only depends on the number of `+' and `-' alleles, thus allowing many equivalent genetic combinations; there are many other stable but suboptimal combinations \citep{Barton:1986vl}. The particular state to which the population converges is thus strongly determined by its previous history. All these suboptimal combinations trap the population in local fitness peaks that deviate considerably from the optimum trait value. We can see in Fig. \ref{Fig:BifurcationZM}A (thick line) that if the effect of each locus on the trait is fairly large, then deviations from the trait that are at most equally large as the effect can maintain any `+' or `-' alleles at equilibrium. Thus, many of the suboptimal combinations are realizable. Also, if the population is resting at an initial equilibrium and the optimum is shifted (either slowly or abruptly), the allele frequencies respond in a coordinated way. Thus, the trait is resilient to perturbations in the sense that all allele frequencies are always equidistant from the bifurcation point where their stability changes, and thus resist large deviations from the optimum. Once the bifurcation point is reached, all loci become unstable at once and suddenly jump to a suboptimal state. Therefore, it is unlikely that populations reach an optimal peak.

In this paper we show that if the loci that constitute the trait have different effects, there is a more heterogeneous distribution of equilibria, with no symmetry amongst peaks. There are still  many  sub-optimal states where the population could get stuck, but we will see that under unequal effects, these suboptimal equilibria are much more similar (and closer) to the optimum. However, the trait is also less resilient to deviations from the optimum, and smaller perturbations render the configurations unstable. In fact, we see in Fig. \ref{Fig:BifurcationZM}A that alleles of very small effects will make $\Delta z / \gamma$ large, implying that the allelic configurations become unstable. Naturally, the occurrence of small effects is contingent on the distribution of allelic effects, which is unknown in detail; we will explore this aspect in this article.

Summarizing, under equal effects precise adaptation to the optimum is harder because the population might get stuck at suboptimal peaks that have large variation and larger mutation load. At equilibrium, selection purges the new mutations, and irrespectively of their allelic effects, each locus contributes  $2\mu/S $ to the genetic variance. In fact, this is an upper-bound achieved when the trait is perfectly adapted to the optimum, irrespective of the distribution of genetic effects (as long as these are larger than their contribution to the genetic variance in equilibrium). However, if the trait mean deviates from the optimum, the genetic variance can differ from that of the HoC \cite{Burger:1994wt} (Fig. \ref{Fig:EqvsUneqDyn}).

A different situation is realized if the allelic effects are smaller than the equilibrium variance, for which the HoC model does not apply. Another classic approximation, which supposes multiple alleles and is often referred as the Gaussian Model (GM) \citep{Kimura:1965ua,Lande:1976wt}, makes the opposite assumption about the allelic effects, i.e. that these are small compared to their contribution to the genetic variance in equilibrium. The GM assumes that there is a continuous production of new alleles that follows a Gaussian distribution of effects at each locus, that is centered at the parental genotypic value. \citet{Barton:1986vl} showed that in polygenic diallelic traits under equal effects, changes in the optimum can lead the population towards stable albeit maladapted equilibria which can have much larger variation than that of the HoC, and fall into a limit that is better approximated by the GM.

The analyses for polygenic systems with unequal effects that we perform here are more challenging than for equal effects. Our current understanding of unequal effects derives from models that deal with a few loci, from which general results are hard to extrapolate \citep{Burger:2001wy,Turelli:1984tb,Chevin:2008ga,Pavlidis:2012ii}. In this article we aim to understand how a trait determined by arbitrarily many loci of unequal effects responds to stabilizing
selection and mutation. Putting aside the technical complexities, we regard this problem as fundamental to understanding the bigger picture of the evolution of polygenic traits, namely that of finite populations subject to drift, and how these are constrained by pleiotropic effects caused by selection on multiple characters. But first, we need to understand in detail the nature of the equilibria and the response of  allele frequencies to  factors such as stabilizing selection and mutation. Thus, we address the simplest case of deterministic selection on a single trait.

We start by studying the equilibria, and find that there can be multiple loci with high polymorphism, provided that they have small effects. For these alleles of small effect, deviations from the optimum trait value are tolerated without affecting their equilibrium. However, we also find that there is a  threshold  $\hat{\gamma} = 2\sqrt{\mu/S}$ that objectively defines which alleles are of ``small'' effect ($\gamma<\hat{\gamma}$), and which are of  {``}large{''} effect ($\gamma>\hat{\gamma}$). The former remain at intermediate frequencies and latter near fixation most of the time. Alleles of large effect can be sensitive even to small deviations from the optimum. In particular, if the optimum is suddenly shifted, we find that the alleles that respond first are those with effects  closer to the threshold value, $\gamma\sim\hat{\gamma}$.
In the long-term, however, the dynamics are intricate. Different initial equilibrium configurations that are equally well adapted may lead the population to totally different regions of the fitness landscape. However, these different genetic states have very similar phenotypic values.

\section*{Model of stabilizing selection and mutation on additive traits}

We consider the simplest diploid genotype-phenotype map, which assumes an additive trait for diallelic loci, without dominance or epistasis:
\begin{equation}
z= \sum_{i=1}^n \gamma_i (X_i + X_i' -1)
\end{equation}
where \(\gamma_i\) is the allelic effect at locus $i$, and $n$ is the number of loci composing the trait, and $X$ and $X'$ are indicators of the `-' allele ($X,X'=0$) or of the `+' allele ($X,X'=1$). We will allow each $\gamma_i$ to vary across loci. Specific values will be drawn from a given distribution (we explore mainly $\Gamma$-distributed effects), although in every run they will be kept constant.
Assuming linkage equilibrium, the trait mean and the genetic variance are given by:
\begin{eqnarray}
\label{Eq:TraitMean}
\bar{z}=\sum_{i=1}^n \gamma _i\left(2p_i-1\right)  \\
\label{Eq:GeneticVariance}
\nu =2\sum_{i=1}^n \gamma _i^2p_i\left(1-p_i\right) 
\end{eqnarray}
where $p_i= \mathcal{E}[X_i]$, the allele frequency of the `+' allele, given by the expectation of $X_i$ in the population. (Unless otherwise stated, the expectations are on the population, not on the distribution of effects).

We assume a Gaussian fitness, \(W_z = \exp[-\frac{S}{2}(z-z_\circ)^2]\) so the mean fitness of the population is:
\begin{equation}
\label{Eq:MeanFitnessSS}
\bar{W}=\exp\left[-\frac{S}{2}\Delta  z^2-\frac{S}{2}\nu \right]
\end{equation}
which assumes weak selection. The genetic load is due to both terms: the deviations from the optimum \(\Delta  z=\bar{z}-z_\circ\) and the genetic variance $\nu $. The maximum mean fitness is 1, which occurs if an optimal genotype is fixed, with no genetic variance.

In an infinite, random mating population, the change in allele frequencies is given by the selection-mutation equation:
\begin{equation}
\label{Eq:AlleleFreqDynamics}
\frac{dp_i}{dt}=-S \gamma _i p_i\left(1-p_i\right)\left(2\Delta   z+\gamma _i \left(1-2p_i\right)\right)+\mu \left(1-2p_i\right) ~,
\end{equation}
for  \(i=1,\ldots, n\), and $\mu,S<<1$ (see for example \citep{Barton:1986vl}). This equation for the dynamics of allele frequencies assumes linkage equilibrium and weak selection.

To understand the complexities of the fitness landscape and how the trait evolves, we first study the properties of the equilibria.  An exact solution of the equilibria of the system defined by Eq. \ref{Eq:AlleleFreqDynamics} for the $n$ loci is possible, but the formulae are complicated, being solutions to coupled cubic equations. Therefore, instead of taking this exact but intricate approach, we will first study the qualitative aspects of the equilibria of Eq. \ref{Eq:AlleleFreqDynamics}, by assuming that $\Delta z$ is given, which uncouples the equations. This will give a solid intuition to understand detailed equilibrium analyses, and how much genetic variance  can be maintained at the different sub-optimal peaks.

Afterwards, we will explore the dynamics and understand the irregular behaviours by using the intuition derived from the equilibrium analyses. We numerically solve the system for all the allele frequencies at each locus. Since we assume diallelic loci, there are $n$ equations to track. We  calculate the genetic variance and trait means from the definitions given by Eqns. \ref{Eq:TraitMean} and \ref{Eq:GeneticVariance}. We typically randomize the initial conditions and the realization of allelic effects, unless otherwise stated.

\section*{Allelic equilibria have two defined regimes}
Equation \ref{Eq:AlleleFreqDynamics} shows that the equilibrium condition for every locus is given by a set of cubic equations coupled through $\Delta z$. When there are many loci,  we can assume a particular value for $\Delta z$ and treat each of the $n$ equations independently. Therefore, for each locus the number of valid roots of the cubic equation depends on four
quantities: the deviation from the optimum $\Delta z$, the allelic effect $\gamma $ of the focal locus, the mutation rate $\mu $, and the
strength of stabilizing selection  $S$. However, these four variables can be combined into just two scaled parameters, \(\delta = \Delta z/\gamma\) and  \(m=\mu /S \gamma^2\) and the equilibrium solution at each locus is given by the scaled equation:
\begin{equation}
\label{Eq:CubicPolynomial}
p^3-p^2 \left(\frac{3}{2}+\delta \right)+p \left(\frac{1}{2}+\delta +m\right)-\frac{m}{2}=0 .
\end{equation}

Figure \ref{Fig:BifurcationZM}A shows how the equilibrium frequencies depend on the scaled deviation from the optimum, $\delta$.
These diagrams also hold for unequal effects, except that the equilibria for each locus are represented by a specific diagram. In Appendix A we give the precise expression of the critical points of Eq. \ref{Eq:CubicPolynomial}.  Figure  \ref{Fig:BifurcationZM}A shows that there may be
two types of equilibria: either near fixation of one or other allele, or a single equilibrium at intermediate frequency 1/2. The factor that determines which equilibrium is attained at a given locus is the scaled variable $m=\mu/S\gamma^2$.

Figure \ref{Fig:BifurcationZM}B shows the equilibrium allele frequencies as a function of $m$. Consider $\delta $=0: we see a partitioning
of two qualitative regions with stable states that are near fixation ($m<1/4$, to the left) and intermediate equilibrium ($m>1/4$,
to the right). 
Notice that since $m$ in inversely proportional to \(\gamma^2\),
 the smaller the effects, the more to the right the alleles are represented in Fig. \ref{Fig:BifurcationZM}B, and {\it vice versa}. Thus \(\hat{\gamma} = 2\sqrt{\mu /S}\) is a threshold that objectively defines alleles of large and small effect: if $\gamma>\hat{\gamma}$, these fall into the category of ``large'' effects, and if $\gamma<\hat{\gamma}$, these fall into the category of ``small'' effects.

Figure \ref{Fig:BifurcationZM}B shows how this diagram is modified for $\delta \neq 0$: the bifurcation is shifted, and the intermediate equilibria close
to $ m\geq 1/4$ are displaced from 1/2. This has two main implications. Firstly, assume a small deviation of $\delta >0$ ($\delta <0$);
some of the alleles of large effect that would have been close to fixation, at the {``}+{''} ({``}-{''}) state, are forced to sweep to the alternative
state. Secondly, some of the alleles of small effect that would be at the intermediate state, $p$=1/2, will show reduced (increased) frequencies.
Most notably, the alleles that are displaced are those that are close to \(m=\hat{m}\), which are those that are close to the threshold \(\hat{\gamma}=2\sqrt{\mu /S}\).

\begin{figure}
\includegraphics[width=\textwidth]{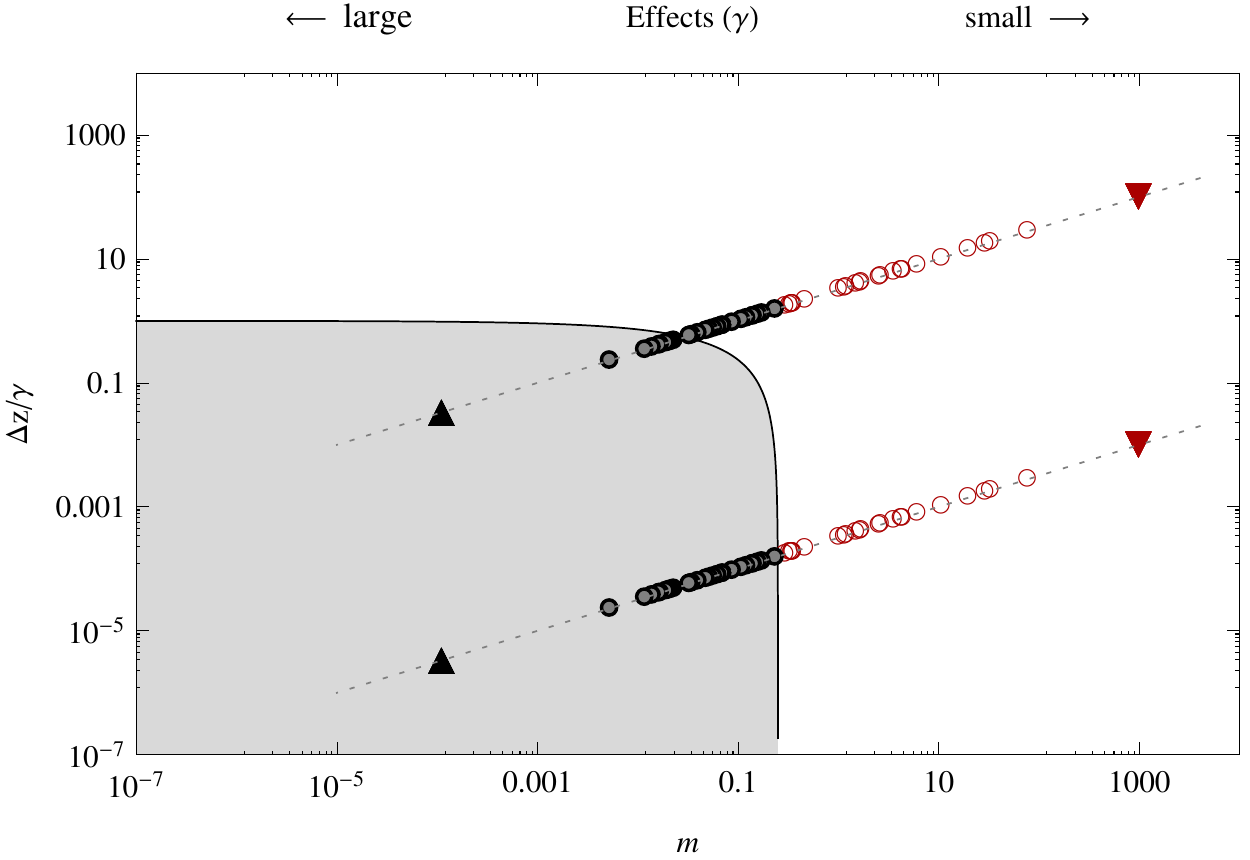}
\caption{Diagram showing the two regions of qualitatively different equilibria of allele frequencies. For $m< 1/4$ (the
shaded region) the allele frequencies are near fixation points. When $m>1/4$ only polymorphisms can be maintained.
On a log scale the effects are distributed along parallel lines whose height is determined by 
$ \log(\Delta z)-\log(\gamma),\log(\mu / S )-2\log(\gamma ) $,  and therefore have a slope of 1/2. Effects that fall at the right hand side of the point $
m=1/4$ can never fall into the bistable regime, and correspond to the alleles with the smallest effect. Depicted in the figure are effects distributed
(bullet) $\gamma  \sim \exp(10)$; (upwards triangle) trait with equal but large effects ($\gamma =3$); (downwards triangle) trait with equal but small effects ($\gamma =0.001$).}
\label{Fig:PhaseDiagram}
\end{figure}

A bigger picture emerges when we consider how the equilibria depend on  combinations of both $\delta$ and $m$ (see Appendix A for the
exact calculations) which is shown in Fig. \ref{Fig:PhaseDiagram}. On a logarithmic scale, the allelic effects fall on a straight line, with the distribution of effects determining their spread along this line:  smaller effects fall towards the right and larger effects falling towards the left of the diagram; different
values of $\delta $ are represented as parallel lines of slope 1/2. 

When effects are large enough that \(m<\hat{m}=1/4\) the alleles can be in a bi-stable regime: there are two stable points close to fixation and one
unstable at intermediate frequency (the thick line in Fig. \ref{Fig:BifurcationZM}A). This is provided that deviations from the optimum are small ($\delta \sim 0$). In this case, the stability is not affected. However (and as we will explain below), deviations that are of the order $\delta ~1/2$ or larger disturb this equilibrium (see Fig. \ref{Fig:BifurcationZM}B).

The situation is very different for small effects, when \(m>1/4\), since there is only one valid root of the cubic above (the thin line in Fig. \ref{Fig:BifurcationZM}A). When the trait is close to the optimum ($\delta \sim $ 0) intermediate frequencies can be maintained, as explained above.
Small deviations from the optimum will readjust frequencies slightly, but the stability of the equilibrium is not modified (there is no qualitative
change in the stability).

 As a consequence, the frequency of alleles of small effect varies smoothly with the deviations from the optimum, whereas those with larger effect experience discontinuous transitions when the magnitude of the deviation approaches half of their respective effects.
 
The scaling properties indicate that the parameters that really matter are $m=\mu/\gamma^2 S$ and $\delta=\Delta z/\gamma$. Thus, the specific numerical choices of $\mu$ and $S$ are not in themselves decisive.
 
\section*{Stability and variation at equilibrium}

\subsection*{Equilibria under mutation-selection balance.}

As above, when mutation is present there might be one, or three solutions for each locus, with the stability depending on the particular multidimensional adaptive peak. Consequently, there might be up to \(3^n\) possible equilibria, although only a fraction of these can be stable. Assuming equal effects, it would be enough to count the number of loci that are fixed or intermediate, since the symmetry of the landscape makes it feasible to understand the stability of any of these
peaks \citep{Barton:1986vl}. With unequal effects, we need to consider each of these configurations separately. These are tractable as long as we assume that $\Delta z=0$, in which case the equilibria at each locus are given by
\begin{equation}
0=\left(\mu -S  \gamma ^2p(1-p) \right)(1-2p) .
\end{equation}
In this case there are three possible states for each locus, namely
\begin{eqnarray}
\label{Eq:NoDevSolutions}
p=1/2 \\
\label{Eq:NoDevSolutionsFixed}
p=\frac{1}{2}\left(1\pm \sqrt{1-4\frac{\mu }{S  \gamma ^2}}\right) ~.
\end{eqnarray}

In Supplementary Information 1 we detail the stability analyses that we now summarize. In the absence of mutation, at most one allele can be maintained polymorphic, irrespective of the magnitudes of the allelic effects and of the deviation from the optimum (this result was anticipated by \citet{Wright:1935vj}). If mutation rates are small compared to selection, the trait mean is exactly at the optimum, and all alleles are of large effects, then the configurations where all loci are close to fixation will be stable (all eigenvalues  are negative). Furthermore, configurations where alleles of large effect are at intermediate frequency will be unstable. We also find that configurations where alleles of small effect are near fixation, are unstable.

Why is this? Alleles with large effects increase the genetic variance substantially. We saw that each allele near fixation contributes $2\mu /S $, whereas if it has intermediate frequency, it contributes \(\gamma^2/2\) to the load. Since the alleles with large effects fulfill that \(\gamma^2/2>2\mu /S\), the genetic load would be much larger if the alleles of large effect were maintained polymorphic. A similar argument applies for allele of small effect. Because \(\gamma^2/2<2\mu /S\), then the load would be significantly higher if these alleles were near fixation. We can interpret this by thinking that the amount of selection required to fix an alleles of small effect would need to be considerably high, as to make them fall into a``large effect" class.

\subsection*{Distribution of allelic equilibria.}
We saw that alleles of large effect will be in near fixation. However, whether they are more likely to be at the `+' or `-' state depends on details such as the position of the optimum and the deviation from it. For example, in Supplementary Information 2 we show that optima positioned towards the largest (smallest) trait value $z_x$ bias alleles to the `+'  (`-') states. Can we estimate how likely are alleles to be close to a particular state?

\begin{figure}
\begin{center}
\includegraphics[scale=0.5]{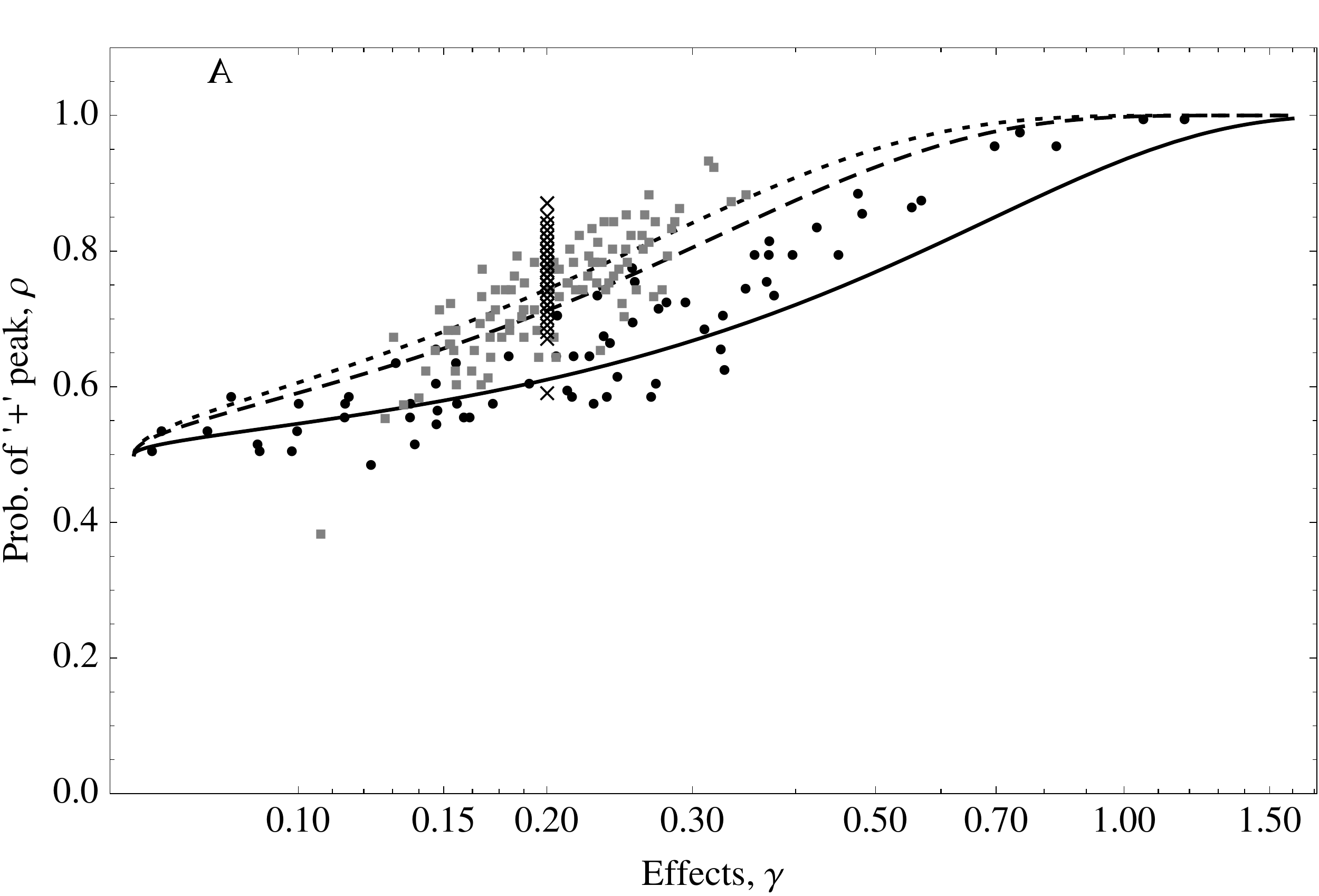}
\includegraphics[scale=0.5]{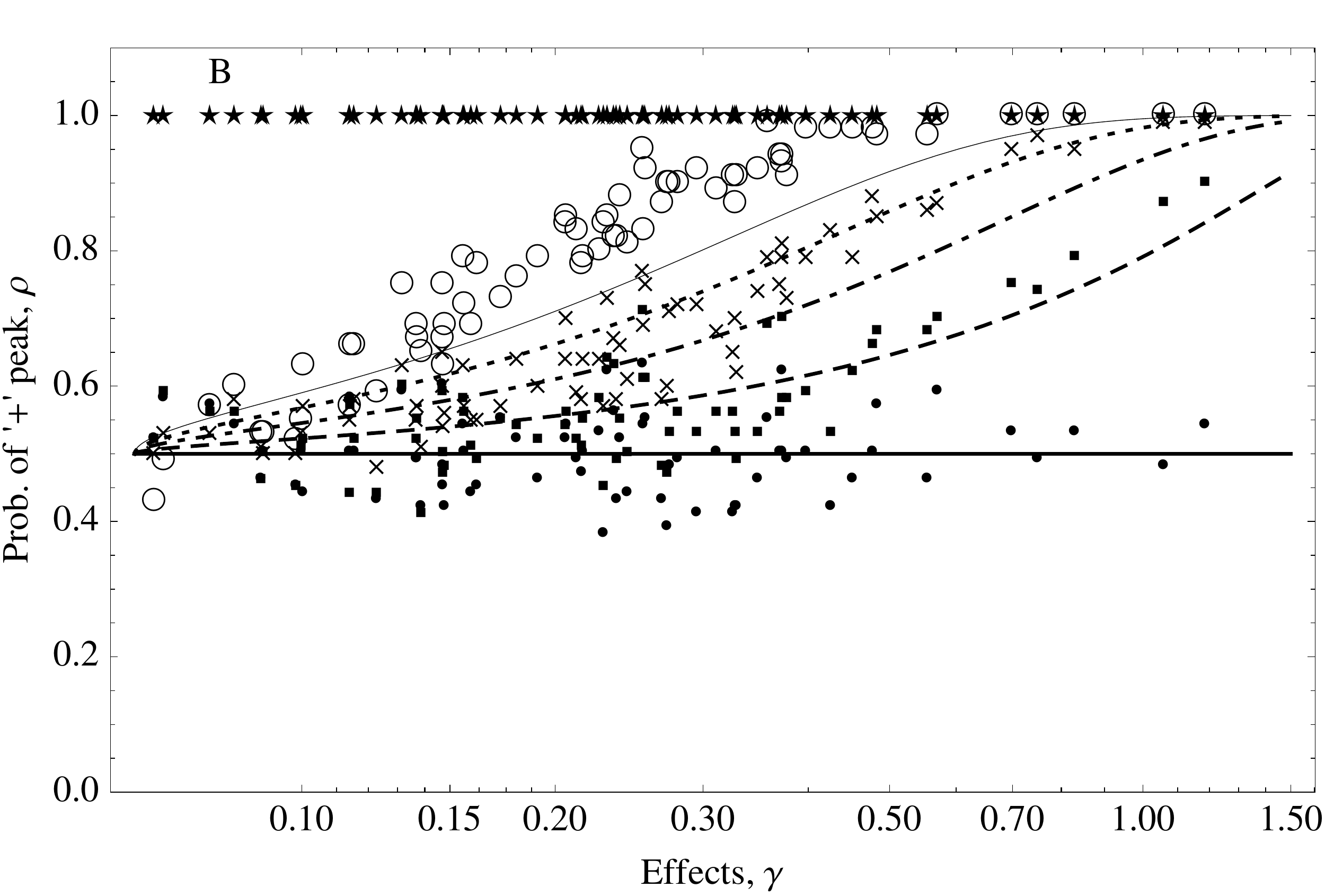}
\end{center}
\caption{The probability of `+' alleles increases as the magnitude of their effect gets larger. Lines: following Eq. \ref{Eq:DistAllelicState}; symbols: average of occurrences of the `+' state from 100 simulations. (A)  The optimum is fixed and the distribution of allelic effects is varied.  $z_\circ=10$ (roughly half-way from the maximum trait value); solid line and bullets: exponential distribution (mean=1/5); dashed line and gray squares: Gamma distribution (shape=20, scale=1/100). Dotted line and crosses: equal effects, $\gamma=1/5$.  (B) The distribution of allelic effects is fixed and the position of the optimum is varied. Effects distributed as an exponential (mean=1/5); thick solid line and bullets $z_\circ=0$, dashed line and squares $z_\circ= 5$, dot-dashed line and crosses $z_\circ= 10$, dotted line and rings $z_\circ= 15$, thin solid line an stars $z_\circ= 20$. In all cases the trait is determined by 100 loci.  $\mu=10^{-4}, S=0.1$. The initial conditions were uniformly and independently drawn for each locus.
}
\label{Fig:PeakProb}
\end{figure}

We  assume that the trait mean is at the optimum, and focus on the state of one particular allele. We will study how the probability $\rho$ that the focal allele  is at the `+' state depends on its effect $\gamma$.  In this case, we take $\rho$ to be a probability calculated over all possible states (peaks), where we assume that the rest of the (background) loci contribute in a way that keep the trait at the optimum. We assume that for all alleles of large effects the initial conditions are such that $\Pr_0(`-')=\Pr_0(`+')=1/2$. Numerically, we perform many runs that start close to uniformly randomly selected peaks, and let the system reach equilibrium. Then we count how often alleles of effect $\gamma$ are in the `+' state. In Appendices B and C we show that that
\begin{equation}
\label{Eq:DistAllelicState}
\rho_j =1\left/\left(1+ \exp\left[-2\frac{z_\circ}{V} \sqrt{\gamma_j^2-4\frac{\mu}{S} }\right] \right) \right. .
\end{equation}
where $V= \sum'_{i\neq j}( \gamma_i^2- \frac{4\mu}{S})$, where $\sum '$  indicates summation on the set of alleles of large effect. Figure \ref{Fig:PeakProb} shows that the predictions of Eq. \ref{Eq:DistAllelicState} are consistent with the simulations. The distribution of effects does not affect the probability of an allele being in the `+' or `-' state, only the effect of the focal allele matters.

The larger the effect of the focal allele is, the larger the probability it is in the `+' state (this assumes positive positioning of the optimum; for negative positioning, the converse would be true). The reason is that alleles that are closer to the threshold value are more prone to the instabilities resulting from small deterministic fluctuations around the optimum. Large alleles, on the other hand, are more often at the `+' state since they are more resilient to perturbations from the optimum value. Thus, once a population attains equilibrium, large alleles with effects close to $\hat{\gamma}$ are much more likely to be stuck in alternative equilibria than larger alleles.

We also find, in Fig. \ref{Fig:PeakProb}B that the larger the value of $z_\circ$, the larger the probability for all loci to be in the `+' peak. This is expected, because lager trait values require more `+' alleles. This obvious observation, although supported by the model, is quantitatively underestimated by it.  In principle, deviations from the optimum trait value can be accommodated in Eq. \ref{Eq:DistAllelicState} (Appendix C). But this correction, at least to first order on $\Delta z$, does not fully account for the underestimation of the model at large optimum values (data not shown). What actually happens is that as the optimum is positioned closer to the range of response of the trait, the distribution of traits is considerable skewed and the Gaussian assumption fails.

Above we saw that alleles with effects that are close to the critical point  are more susceptible to small perturbations to the optimum. Thus how stable the trait is to small shifts of the optimum, depends on how far the frequencies are from the critical point, which in turn depends on the particular distribution of allelic effects. Under equal effects, all alleles are equally far from the critical point, and thus remain stable for a long period until the deviation is large enough. But when the deviation reaches the critical value all alleles are perturbed at the same instant.

Under unequal effects the picture is more complex. The individual equilibria of each allele are perturbed differently by deviations from the optimum. Moreover, once a given allele is perturbed and placed at an alternative state, the newly attained equilibrium is characterized by a different deviation from the optimum, potentially perturbing yet another allele. The interplay amongst the complex equilibria are hard to characterize in detail.

Now we will determine the size of the deviations from the optimum. By employing perturbation analysis (Appendix B) we find that positive deviations from the optimum push allele frequencies closer to fixation. We also prove that the maximum deviation close to a given peak is of the order $\tilde{\Delta}z \simeq \min_{i\in \mathbb{L}} \gamma_i/2$, where $\mathbb{L}$ is the set of large effects (in Appendix B we give an exact expression to the maximum deviation). Clearly, this is bounded below by $\hat{\gamma}/2$, and $\tilde{\Delta}z$ depends on the particular draw of effects. This limit for the deviation is suggested by the diagram Fig. \ref{Fig:BifurcationZM}A:  we see that the shoulders of the black lines actually occur relatively near $\delta=1/2$ (as long as effects are large enough). Consequently, we expect that most of the time the traits will be fairly well adapted, and most of the load is given by the genetic variance, rather than by large deviations of the mean from the optimum.

\subsection*{Genetic variance.}
By direct substitution of Eqns. \ref{Eq:NoDevSolutions}-\ref{Eq:NoDevSolutionsFixed} into Eq. \ref{Eq:GeneticVariance} (see Table 1) we see that the genetic variance that is maintained by mutation-selection balance is \(\gamma^2/2\) per locus at the intermediate state, and \(2 \mu/S\) per locus near fixation. Contrast this to the genetic variance predicted by the HoC, which is same as for traits controlled by equal but large effects, i.e. $\nu = 2 n \mu/S $. Under unequal effects , if $\Delta z=0$, the genetic variance is
\begin{equation}
\label{Eq:GeneticVarianceMain}
\nu =2n_f\frac{\mu }{S}+\frac{1}{2}\sum _{k\in \mathbb{S}} \gamma_k^2
\end{equation}
where $n_f$ is the number of alleles of large effect, and the set $\mathbb{S}$ contains the $n_s$ alleles with small effects, \(\gamma^2/2<2\mu/S\); clearly, $n=n_f+n_s$. Notice that the first term is due to alleles that are close to fixation, and their contribution to the genetic variance is independent of their effect, and the second term is due to alleles of small effect, which are at intermediate frequency. Equation \ref{Eq:GeneticVarianceMain} is one of our central results. 

With this result we come back to Fig. \ref{Fig:EqvsUneqDyn}: in panel B the genetic variance of the trait with unequal effects is lower than the HoC. That is because  24 alleles are of small effect.  Equation \ref{Eq:GeneticVarianceMain} correctly predicts the  equilibrium variance, $\nu = 0.064$. However, note that if we use the HoC with $n_f$ (instead of $n$) then $\nu \simeq 0.052$,  more than 80\% of the total variance.

Thus, the HoC variance bounds the genetic variance under unequal effects. Specifically, Eq. \ref{Eq:GeneticVarianceMain} implies that \( \nu_{HoC}(n_f) \leq \nu \leq \nu_{Hoc}(n)\), where $\nu_{HoC}(m)$ is the HoC variance with $m$ loci. With no deviations from the optimum, the load is proportional to the genetic variance. Under the HoC the load is always $L=S\nu/2 = n\mu$. However because under unequal effects the genetic variance is smaller, the mutational load will also be smaller, and dependent on the distribution of alleles. 

\begin{figure}
\begin{center}
\includegraphics[scale=0.5]{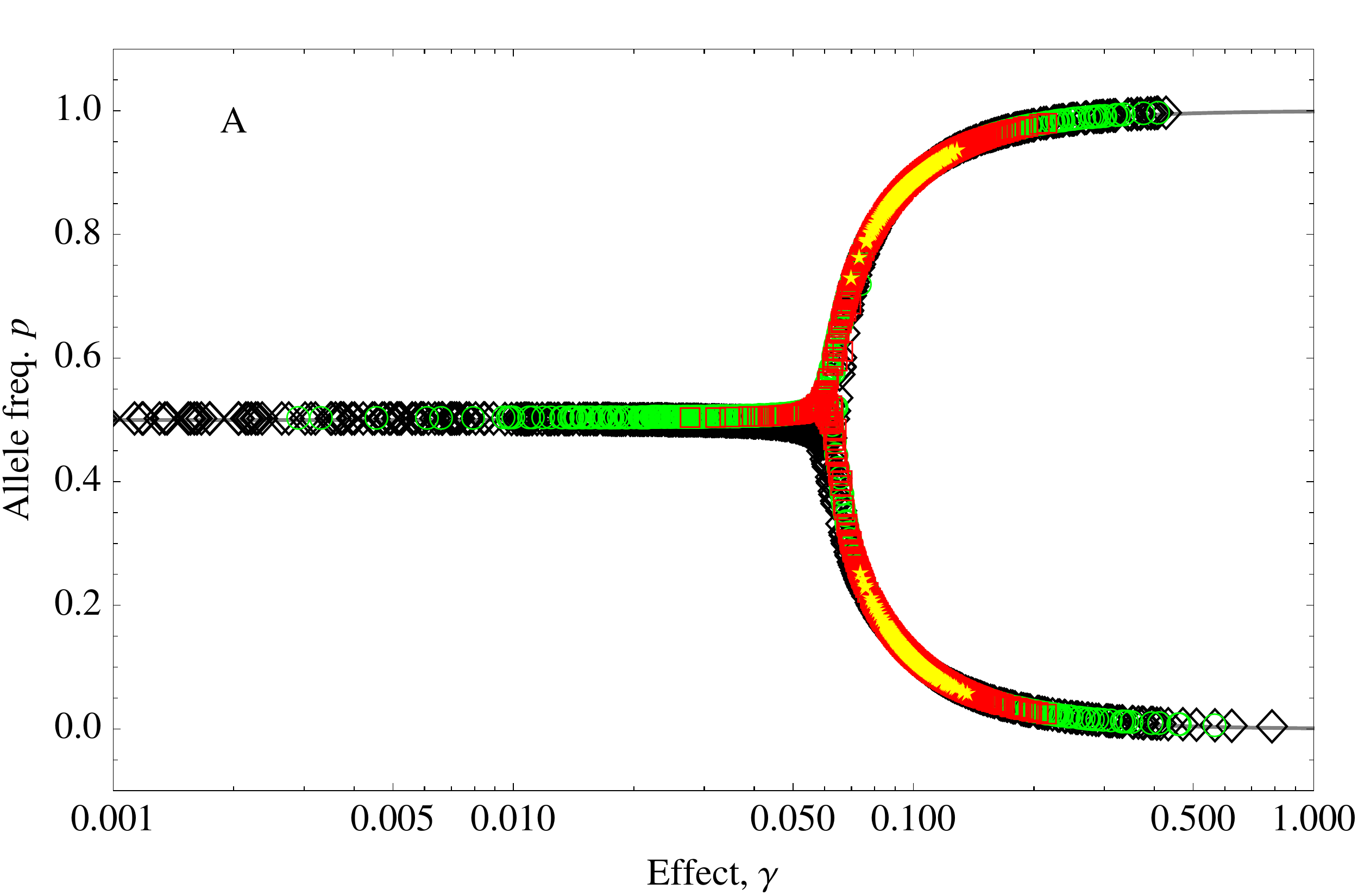}
\includegraphics[scale=0.5]{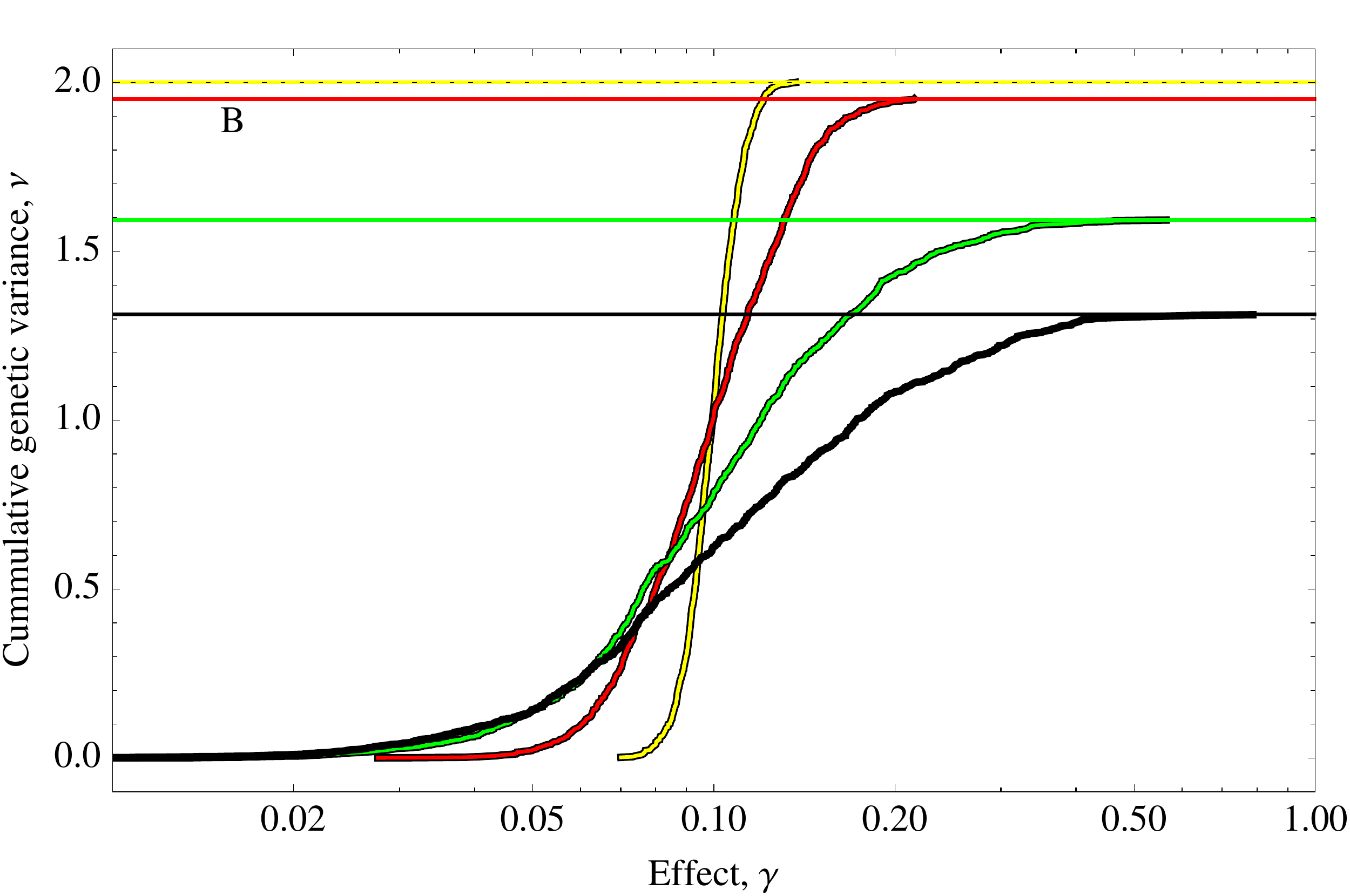}
\end{center}
\caption{(A) Equilibria under different distributions of allelic effects. Symbols: data from simulations. Initial frequencies were drawn uniformly in $(0,1)$ and the system numerically evolved to equilibrium. Gray lines: equilibria of allele frequency; symbols: numerical equilibria. 
(B) Cumulative contribution to the genetic variance under different distributions of allelic effects $\mathcal{P}(\gamma)$. Solid curves: data corresponding to the simulations in panel A. Solid horizontal lines: equilibrium genetic variance (Eq. \ref{Eq:GeneticVarianceMain}). Dotted line: genetic variance of the HoC.  $\mathcal{P}(\gamma)$ is a Gamma distribution of mean=0.1 with shape parameters $\pi$ as: yellow, stars: $\pi=10^{-3}$, red, squares $\pi=10^{-2}$, green, circles $\pi=5 \times 10^{-2}$, black, diamonds $\pi=10^{-1}$. $\mu=10^{-4}, S=0.1,n=1000, z_\circ=0$.}
\label{Fig:DistEffects}
\label{Fig:GenVarDistribution}
\end{figure}

The equilibrium genetic variance depends on the distribution $\mathcal{P}(\gamma)$ of allelic effects. Even though alleles near fixation contribute to $\nu$ independently of $\gamma$, the proportion of alleles of large effects will change with $\mathcal{P}$. For example, fixing the expected value of $\gamma$ at a value larger than $\hat{\gamma}$, but allowing the shape of the distribution to change, will result in different proportions $P=n_f/n$ of alleles of large effect (Fig. \ref{Fig:DistEffects}). In this way, we keep the whole range of response of the trait comparable across different distributions of effects. Distributions peaked around the mean will correspond to traits with alleles of large effect, all of which will be near fixation (Fig. \ref{Fig:DistEffects}A, yellow stars), thus 100\% of the variance is due to alleles of large effects and will match the HoC variance  (Fig. \ref{Fig:GenVarDistribution}B, yellow curve). For distributions that are more spread, the traits will have mixed effects (Fig. \ref{Fig:DistEffects} red squares/line and green circles/line): $n_s$ increases to 96, with about 7\% of the variance due to alleles of small effects, and $n_s=342$, about 17\% of the variance is due to alleles of small effects, red and green respectively. The extreme case will be for positively skewed distributions, such as the exponential, where the proportion of alleles of large effects is much smaller (Fig. \ref{Fig:DistEffects} black diamonds), and the genetic variance will be considerably lower than that of the the HoC (Fig. \ref{Fig:GenVarDistribution}B, black curve): roughly half of the alleles ($n_s=478$) are of small effect, but contribute by 20\% to the total variance.

\subsection*{Distribution of phenotypic equilibria}

\begin{figure}
\begin{centering}
\includegraphics[width=\textwidth]{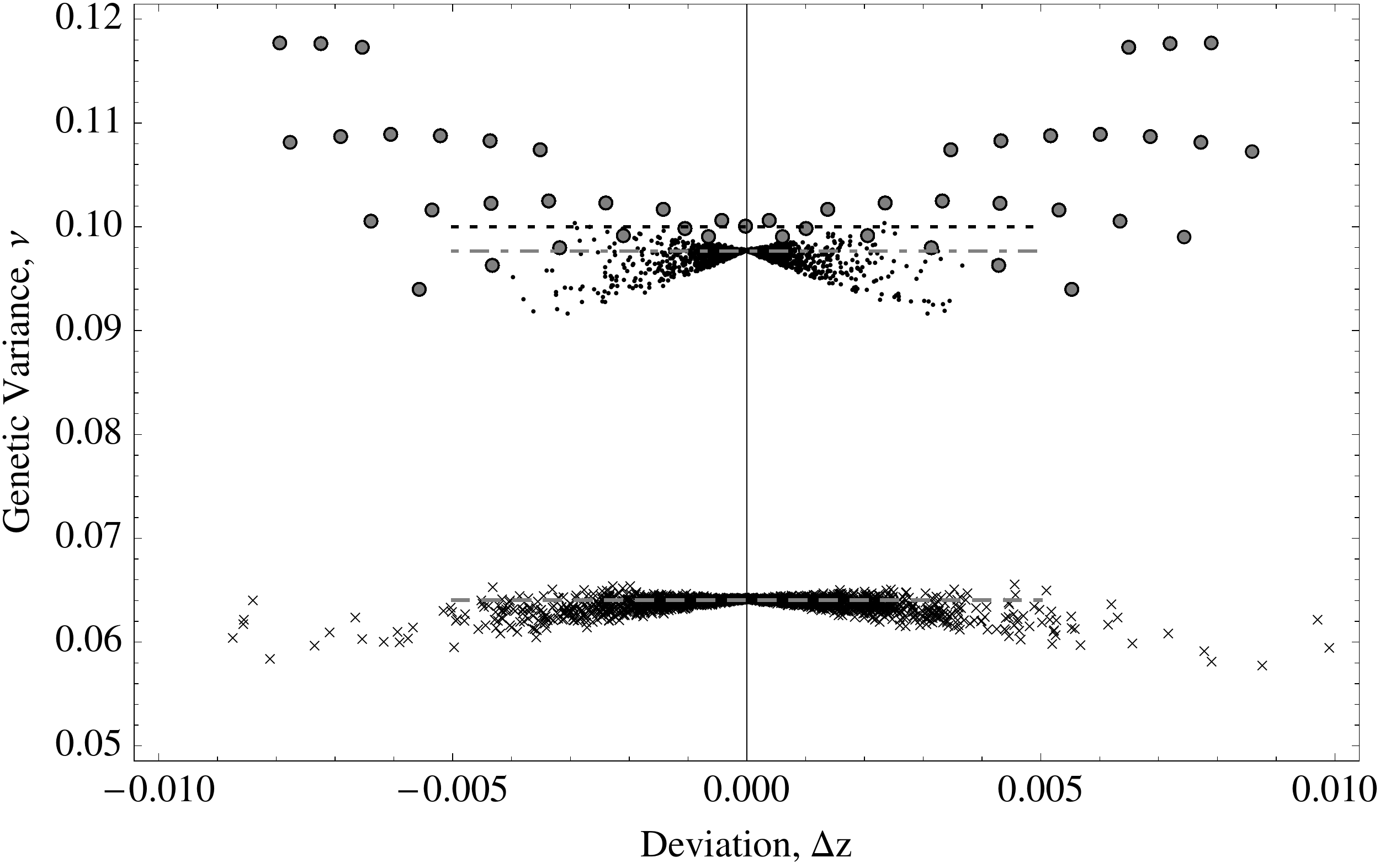}
\end{centering}
\caption{Phenotypic equilibria under different distribution of effects. Gray circles: equal effects; dots: effects tightly clustered around the mean (gamma-distributed with variance=1/1000); crosses: exponentially-distributed effects (variance=1/100). In all cases the mean effect is 0.1.Points are results from numerical calculations for 11 equidistant optima $z_\circ \in [-z_x/2,z_x/2], z_x=\bar{\gamma}=5$, at each point employing 200 runs with uniform random initial conditions. $\mu=10^{-4}, S=0.1,n=50$.}
\label{Fig:QuantitativeEquilibria}
\end{figure}

For many loci, the number of allelic equilibria can be  astronomical. Nevertheless, under equal effects it can be calculated explicitly \citep{Barton:1986vl}. When mapped to trait mean and genetic variance, the number of distinct equilibria is smaller, since many combinations of allelic effects have equivalent, or at least very similar, trait mean and variance. Fig. \ref{Fig:QuantitativeEquilibria} shows how the phenotypic states change when we keep the mean of the allelic effects constant, but increase its variance: the genetic variance decreases (see also Fig. \ref{Fig:GenVarDistribution}), and deviations from the optimum have less effect on the genetic variance, making Eq. \ref{Eq:GeneticVarianceMain} a good approximation. Notably, we find that the number of values of trait mean and genetic variance increase when the distribution of effects spreads. However, these equilibria become more similar and closer to each other.

\begin{figure}
\begin{center}
\includegraphics[scale=0.3]{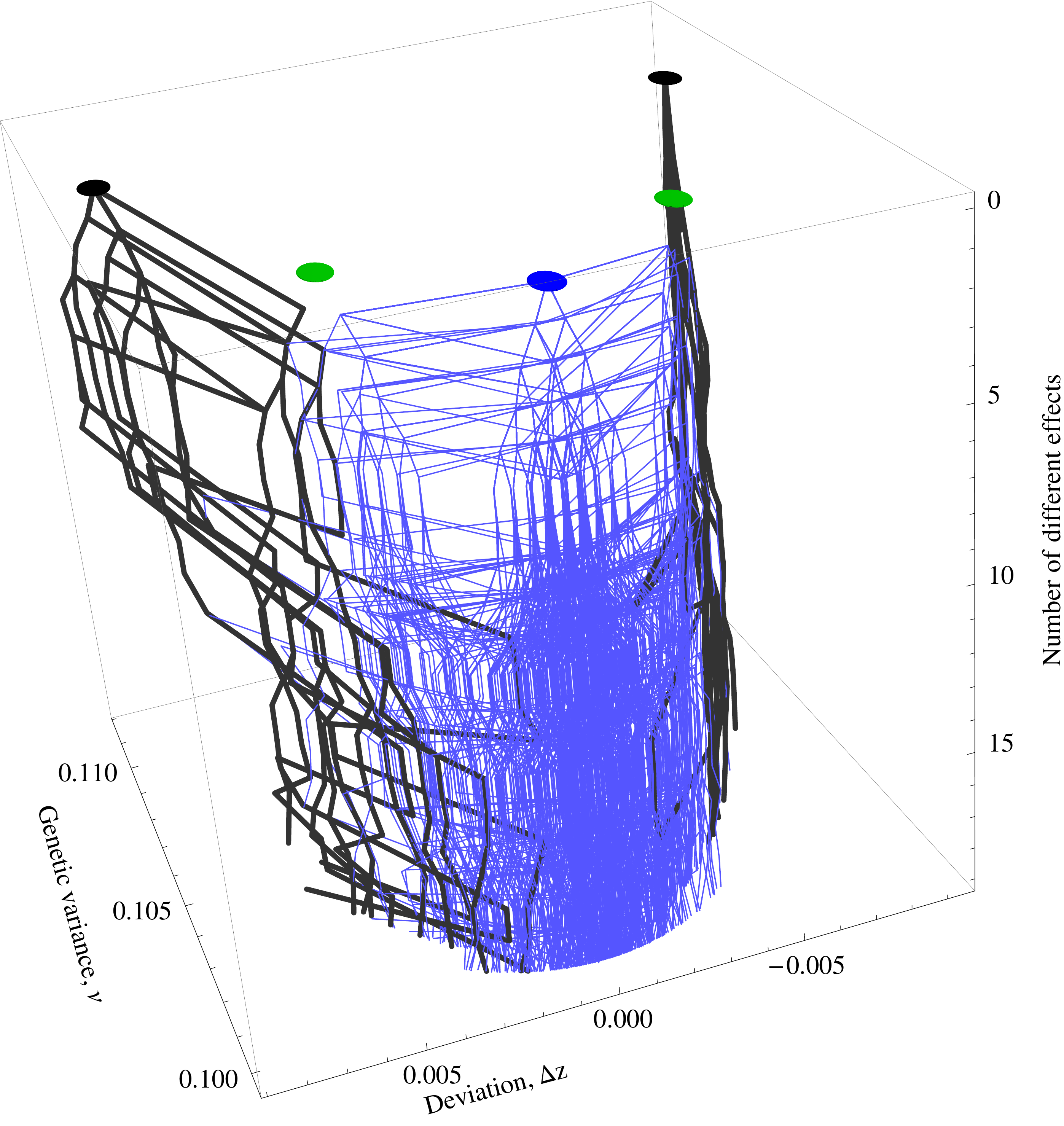}
\includegraphics[scale=0.5]{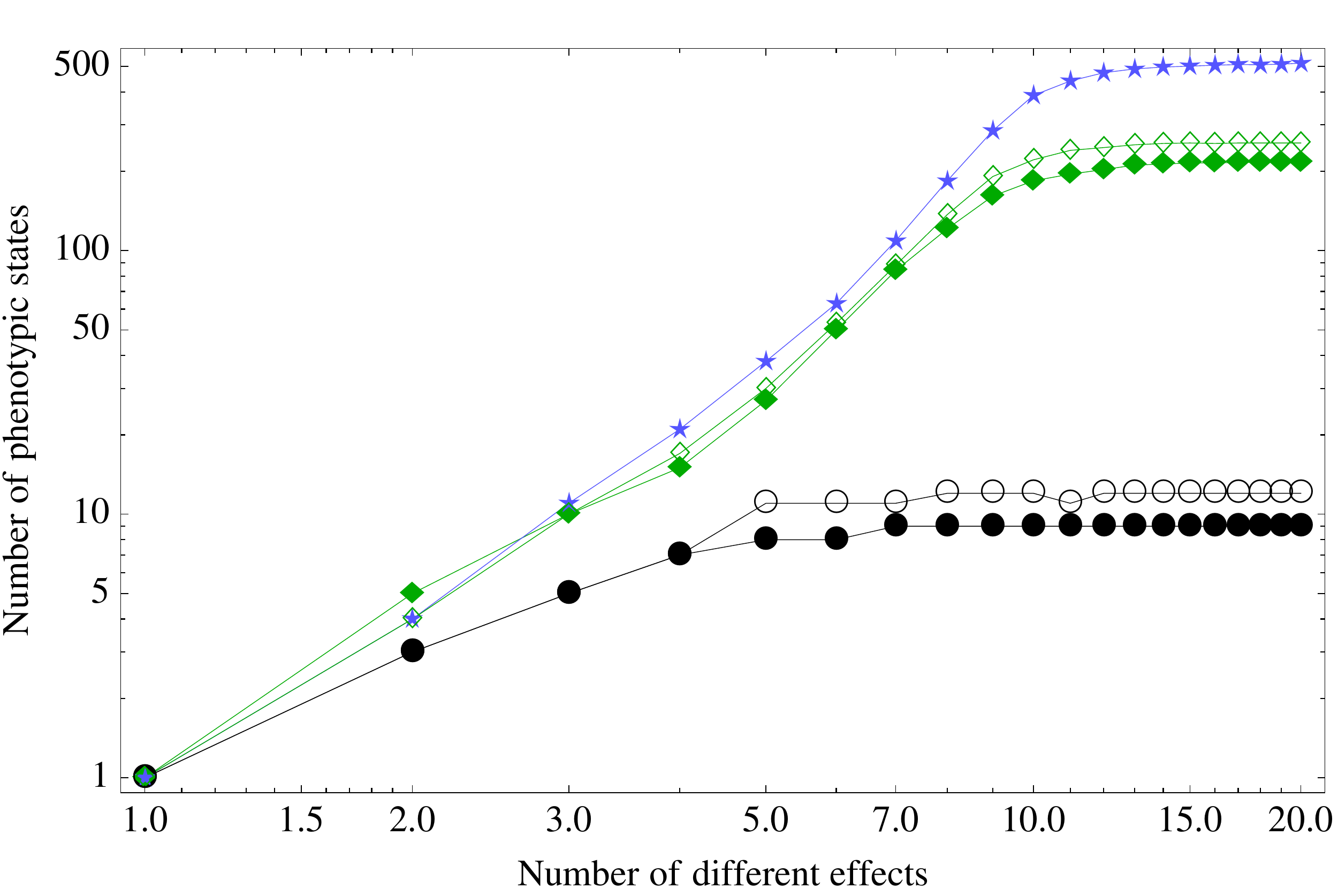}
\end{center}
\caption{
(A) Connectedness of macroscopic fitness peaks as the number of different effects of a polygenic trait are systematically increased from 1 to 20. The black thick lines originate at sub-optimal equilibria, and the thin blue lines (color online) originate at the optimal equilibrium ($\delta z=0$).
(B) Number of new equilibria derived from the sub-optimal equilibria under equal effects (black circles and green diamonds) and at the optimum equilibria (blue stars) as the number of different effects is increased. $\mu=10^{-4}, S=0.1,n=50, z_\circ=0$.}
\label{Fig:LandscapeComplexity}
\end{figure}

Although it is hard to count the states precisely, we can study how a given phenotypic equilibrium is affected as the asymmetry of the unequal effects is increased. For instance, suppose under equal effects we track a set of initial conditions $\Pi$ that lead to a particular point in the ``phenotypic'' $(\bar{z},\nu)$-space. The states to which these trajectories converge (basins of attraction) are symmetric in the sense that exchanging `+' alleles at one locus with `-' alleles at another does not change the phenotypic states. If we keep constant the set $\Pi$, but now change one effect by a small amount, how is the distribution of phenotypic states affected? Many of the trajectories starting at $\Pi$ that under equal effects converged to equilibria characterized by the same trait mean and variance will now converge to different points in $(\bar{z},\nu)$-space. The symmetries are broken and exchanging alleles at that locus with another one affect the trait mean and variance. Hence, the phenotypic equilibria show bifurcations when we perturb the effects. We can repeat this procedure by then perturbing a second locus, and so on. Thus, if we represent the phenotypic states as nodes, and we connect these nodes according to the initial condition that led to their corresponding phenotypic states, we will have a graph that represents the increase in complexity of the adaptive peaks. Unless we have a way to cover the initial space uniformly, this does not ensure complete counting of the number of phenotypic equilibria. However, even if the subsampling of the basins of attraction is poor, the method quantifies how complex the space becomes as we increase the variance of the allelic effects. 

If we carry out this procedure for different phenotypic states under equal effects, then we will have several of these graphs. If these graphs share nodes, then it means that the adaptive landscape is more accessible to better-adapted equilibria (because under unequal effects the equilibria have less genetic load, Fig. \ref{Fig:QuantitativeEquilibria}). In fact, in Fig. \ref{Fig:LandscapeComplexity} we see an example for three graphs derived from the optimal peak and two sub-optimal peaks.

Altogether, this exposes that under unequal effects the fitness landscape is more complex or ``rough", but the solutions are generally closer to the optimum. Surprisingly,  perturbing slightly the effect of only one or two loci is enough to overlap different graphs with common nodes, indicating that unequal effects act like a funnel to guide the trajectories to nearly optimal states.

\section*{Initial response to selection}

We saw that there are two well-defined regimes which clearly separate alleles of large effect from alleles of small effect. If the optimum shifts,
which alleles respond first? A related question is: is the initial rate of change of the trait driven mainly by alleles of large or small effect?
Although these two questions are related, they are not the same; even if, for example alleles of small effect sweep first, they might not drive a
substantial displacement of the trait. Conversely, even if an allele of large effect sweeps first, its overall effect might be negligible when compared
to a background of very many loci of small effect. 

To calculate the rate of response of an allele, assume that the population is at equilibrium at a local peak with no deviation
from the optimum that is at \(z_\circ\). Suddenly, the optimum is placed at another value \(z_f\). Equation \ref{Eq:AlleleFreqDynamics} implies that at each locus
\begin{equation}
\label{Eq:AFoptimumShift}
\frac{dp}{dt}=2S  \gamma   p  q \Delta \Omega ~,
\end{equation}
where \(\Delta \Omega =z_f-z_\circ\). For alleles of small effect, the right hand side is \(\frac{S \gamma }{2}\Delta \Omega\). Therefore, alleles with infinitesimally small effects will be nearly neutral and will  have a vanishingly small rate of response to selection. As the effects become closer to (but still smaller than) $\hat{\gamma}$, the rate of response is larger. Consider now alleles that have infinitely large effects. The right hand side of Eq. \ref{Eq:AFoptimumShift} is \(\frac{2 \mu }{\gamma }\Delta \Omega\), and implies that  since these alleles will be almost fixed, there is little variation to select on. Consequently, their rate of response is also vanishingly small. As the effect become closer to (but still larger than) $\hat{\gamma}$, the rate of response become larger. Therefore, those alleles with effects close to $\hat{\gamma}$ will have the earliest response to selection because they are the most sensitive to deviations from the optimum. Thus, the maximum response for each limit is given by the effect that is exactly at the critical value\(\hat{\gamma}\). Evaluating Eq. \ref{Eq:AFoptimumShift} at \(\hat{\gamma }\) we find that the rate of response is at most  
\begin{equation}
\label{Eq:MaxResponse}
\left(\frac{dp}{dt}\right)_{\max }=\sqrt{\mu   S}\Delta \Omega  ~,
\end{equation}
indicating that alleles with effects close to \(\hat{\gamma}\) drive the initial response of the trait to selection.

\section*{Long-term response to selection}

The long-term response of a polygenic character to the displacement of the optimum trait value can be driven by alleles other than those of intermediate effect. Although the alleles of large effect evolve slowly in the beginning, they can eventually gain representation and evolve much faster. A general closed solution for the dynamics is neither possible nor useful, as the behaviour of the allele frequencies is rather complicated. The question is whether the theory developed above can be useful to gain insight into the long-term response of the trait.

\subsection{Abrupt displacement of the optimum.}

We will assume that re-positioning the optimum happens always within the range of response of the trait, and far from the extreme values given by \(z_x=\sum \gamma _i\);  i.e. \(-z_x<<z_\circ<<z_x\). The equilibrium analyses revealed that the particular position of the
optimum is not decisive for equilibria or stability. Instead the deviation from the optimum is the important
factor. Thus, if the population eventually adapts to the new optimum value, the genetic variance that is maintained at the newly established equilibrium will be more or less the same  as in the beginning. In the transient time, the dynamics will be complicated and depend on the specific initial conditions (the adaptive peaks where the population initially stands), and on the distance to the optimum.

\begin{figure}
\includegraphics[width=\textwidth]{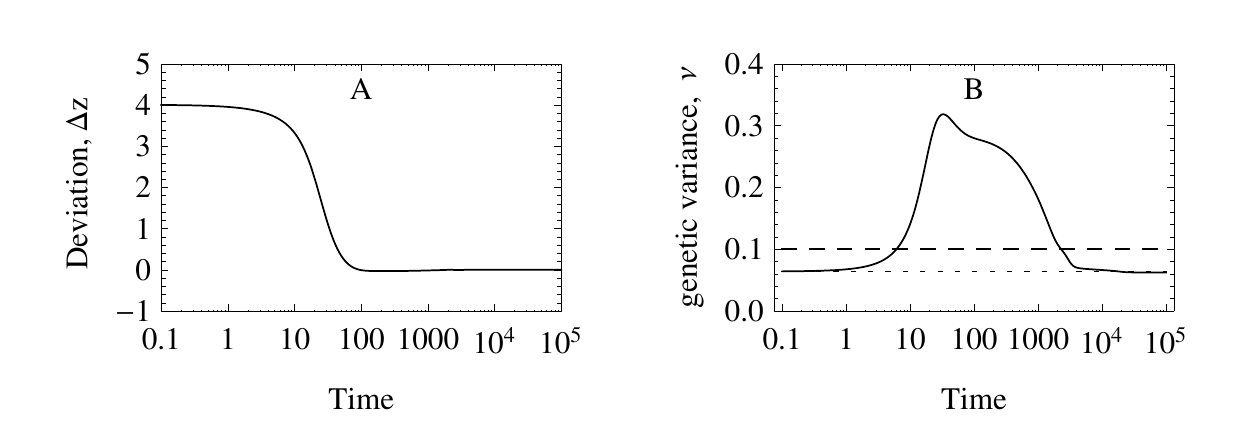}
\includegraphics[width=\textwidth]{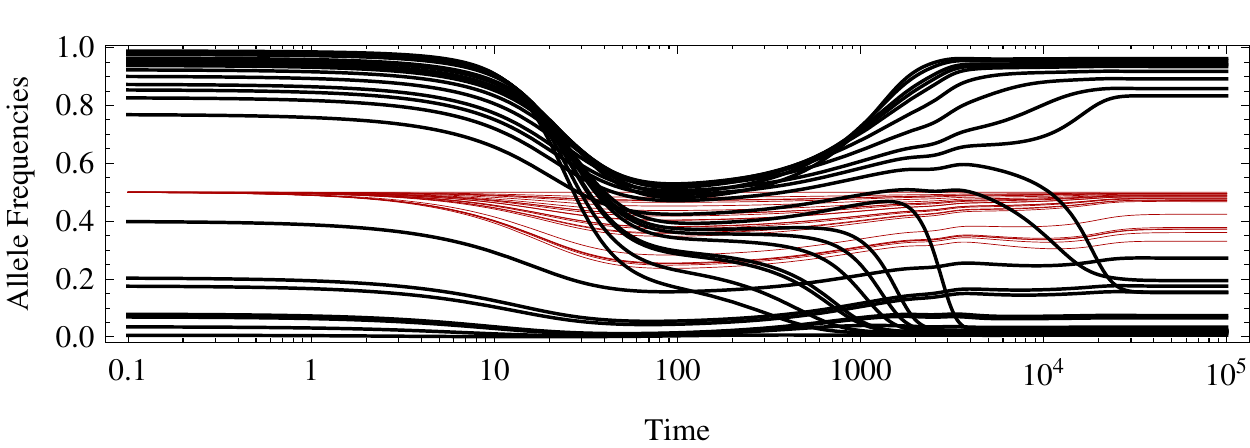}
\caption{Response to an abrupt displacement of the optimum of a polygenic trait. (A) Deviation of the trait mean from the newly positioned optimum. (B) Genetic variance. Black: exact numerical results. Dashed black line: house of cards prediction,  ($\nu = 2 n\mu/S$). Dotted black line: exact value from Eq. \ref{Eq:GeneticVarianceMain}. (C) Response of the allele frequencies; black lines: alleles of large effect; thin red lines (color online): alleles of small effect. The trait is constituted by $n=50$ loci; 26 of large effects, and 24 of small effects, distributed as an exponential of mean $=1/10$. $\mu=10^{-4}, S=10^{-1}, z_\circ \simeq -2$.}
\label{Fig:OptimumShiftZV}
\end{figure}

If the optimum changes abruptly and is larger in magnitude than the largest of the effects, all the equilibria will be perturbed, favouring an increase in the frequency of the  those alleles that diminish the deviation from the optimum. That is, if the new optimum value is smaller (larger) than that of the original optimum value,  `-' (`+') alleles will increase in frequency. This displacement is seen in the diagram of Fig. \ref{Fig:PhaseDiagram}  towards the top (where only one stable border is initially beneficial), and then a gradual return of the line to low values of $\Delta z$. Therefore, we expect to observe a transient
increase in the genetic variance. Fig. \ref{Fig:OptimumShiftZV} provides an example where these patterns are in fact found.

In the example of Fig. \ref{Fig:OptimumShiftZV}, of 50 loci, 26 are of large effect and contribute more than 80$\%$ of the initial variance, whereas 24 are of small effect, contributing the remaining 20$\%$ of the variation. In Fig. \ref{Fig:OptimumShiftZV}C we see that many of the large alleles shift in frequency and some sweep, transiently raising the genetic variance. In this case, since the optimum shifters from $z_\circ=2$ to  $z_\circ=-2$, the `+' (`-') alleles decrease (increase) in frequency. Alleles of small effect are displaced, but not substantially. Most of the transient variation that is generated is due to sweeps of alleles of large effects.

As hypothesized above, even if transient dynamics are very complex (Fig. \ref{Fig:OptimumShiftZV}), when a new equilibrium is attained, the final deviations from the optimum are small, and the genetic variance is close to that of Eq. \ref{Eq:GeneticVarianceMain} (see also Fig. \ref{Fig:EqvsUneqDyn}). As we saw in the introduction, under equal effects the population will evolve to a sub-optimal state where plenty of genetic variation is maintained. Why do populations end up better adapted under unequal effects? 

At first we might think that a bulk of alleles of small effects could provide enough background variation, allowing the population to explore the genetic space more efficiently. However, the response of a trait constituted only by alleles of large effect is virtually the same as that of a trait that contains also alleles of small effect, in as much as the initial genetic variation contributed by the latter alleles is small, and the optimum is not too close to the maximum trait value (Supplementary Information 3). Thus, concerning the response, alleles of small effect can be regarded as nearly neutral.

Another plausible explanation lies in the rate of ``beneficial mutations'' (in the sense that these are mutations that approach the trait mean to the new optimum). Because under equal effects the response of allele frequencies is synchronized, most mutations are initially beneficial. However, due to the epistatic nature of the fitness landscape, those initially beneficial mutations are not necessarily beneficial once the rarer alleles have increased their representation in the population, and might even become detrimental. Furthermore, these alleles arise and increase their frequency on the same time scale. Under unequal effects different mutations arise at different times and can compensate the load contributed by previous mutations. In fact, because of epistasis, we expect and  in fact we find (Fig. \ref{Fig:OptimumShiftZV}C), that some alleles that are initially beneficial increase in frequency, but afterwards, become detrimental and decrease in frequency again. This ``prevention of sweeps'' has been observed in polygenic traits with up to 8 loci \citep{Pavlidis:2012ii}. However, under equal effects allele frequencies remain synchronized along evolution, and it is unlikely that the initial conditions and the shift in the optimum are in general finely tuned in such a way to allow the population to reach a local peak that is close to the global optimum.

Since the previous examples suggest that adaptation is driven mostly by alleles of large effect, an interesting question that follows is: what happens  when traits are controlled principally by alleles of small effect? First of all, Eq. \ref{Eq:GeneticVarianceMain} indicates that the genetic variance will be much lower than that of the HoC. We find that the population eventually adapts (although somewhat slower), but with virtually no change in the genetic variance. In this example, the trait has only three alleles of large effect, which contribute by 11\% of the genetic variance, and by 397 loci with alleles of small effect, which contribute the remaining 89 \% of the variation. The three alleles of large effect sweep, but ultimately don't affect the genetic variation substantially (assuming their HoC contribution), and some alleles of small effect are strongly shifted.

\begin{figure}
\includegraphics[width=\textwidth]{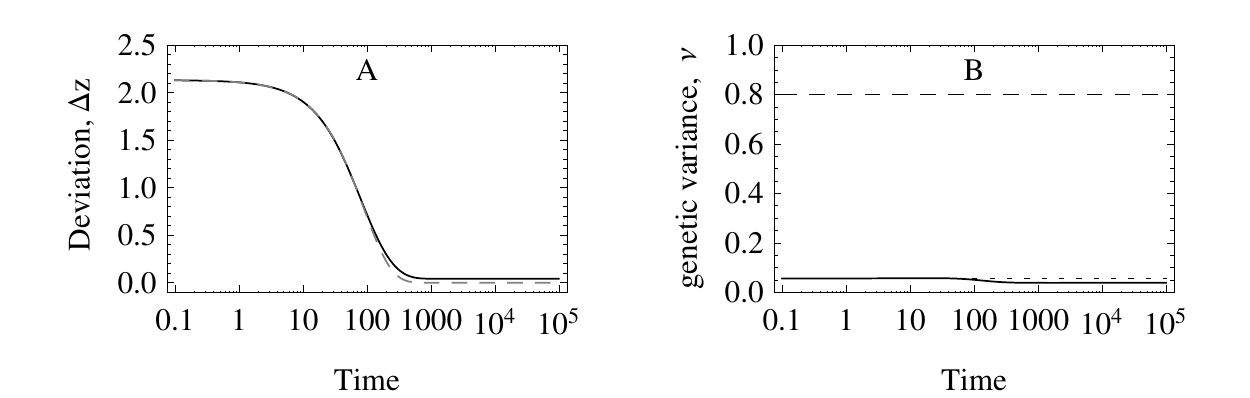}
\includegraphics[width=\textwidth]{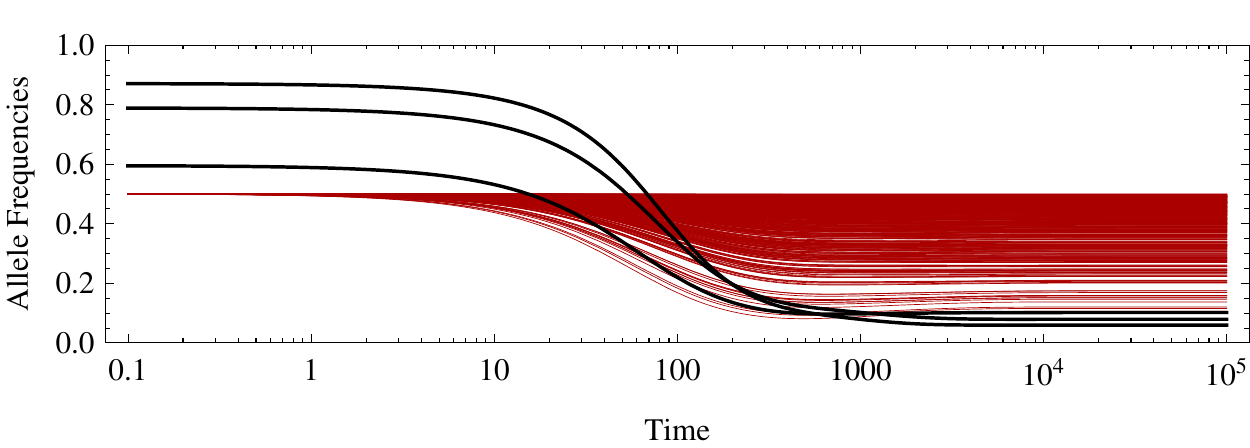}
\caption{Response to an abrupt displacement of the optimum of a polygenic trait constituted mainly by alleles of small effect. (A) Deviation of the trait mean from the newly positioned optimum. Black line: exact numerical results; Dashed gray line: approximation assuming an effective number of loci, $n_{eff}=28$ and constant genetic variance, $\nu \simeq 2 n_e\mu/S$ (see text). (B) Genetic variance. Black: exact numerical results. Dashed black line: house of cards prediction,  ($\nu = 2 n\mu/S$). Dotted black line: exact value from Eq. \ref{Eq:GeneticVarianceMain}. (C) Response of the allele frequencies; black lines: alleles of large effect; thin red lines (color online): alleles of small effect. The trait has $n=400$ loci; 3 of large effects, and 378 of small effects, distributed as an exponential of mean $=1/80$. $\mu=10^{-4}, S=10^{-1}, z_\circ \simeq -2$.}
\label{Fig:ManySmallAllelesZV}
\end{figure}

Figure \ref{Fig:ManySmallAllelesZV} shows the response of a system of alleles of principally small effects. Although 400 loci determine the trait, we estimate that the effective number of alleles $n_e\simeq28$ (see Supplementary Information 4). Assuming constant genetic variance given by the HoC (but using $n_e$) , we find that directional selection towards the optimum explains the response of the trait (although it fails to predict a minor final deviation from the optimum). This experiment highlights why we might find stasis of the genetic variance and a sustained response to selection, which is caused  by innumerable alleles of small effect. Under these circumstances, although experimental essays would be able to detect only a few major loci \citep{Hindorff:2009cc,Visscher:2012je}, these turn out to be the least relevant to explain the quantitative genetic variation.

\subsection{Slowly moving optimum.}

\begin{figure}
\includegraphics[width=\textwidth]{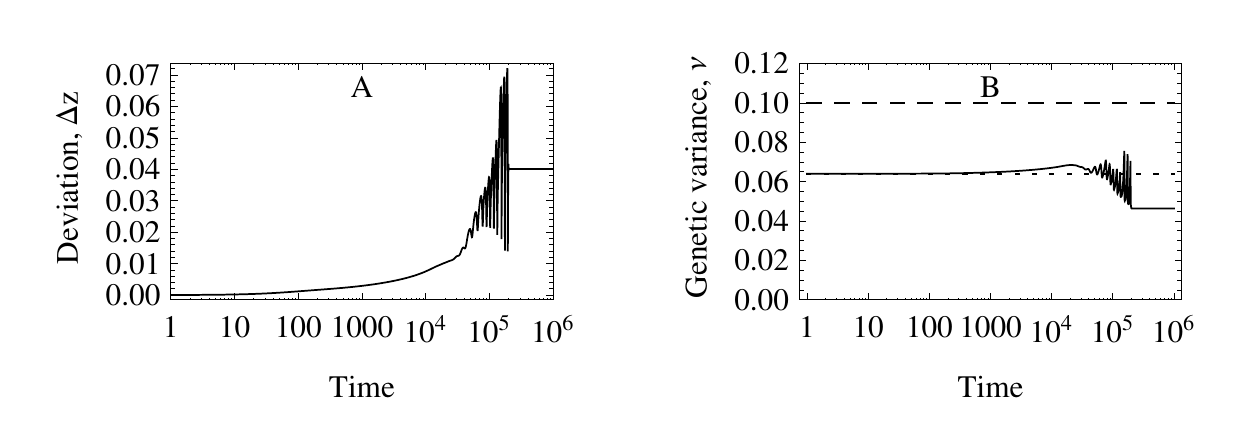}
\includegraphics[width=\textwidth]{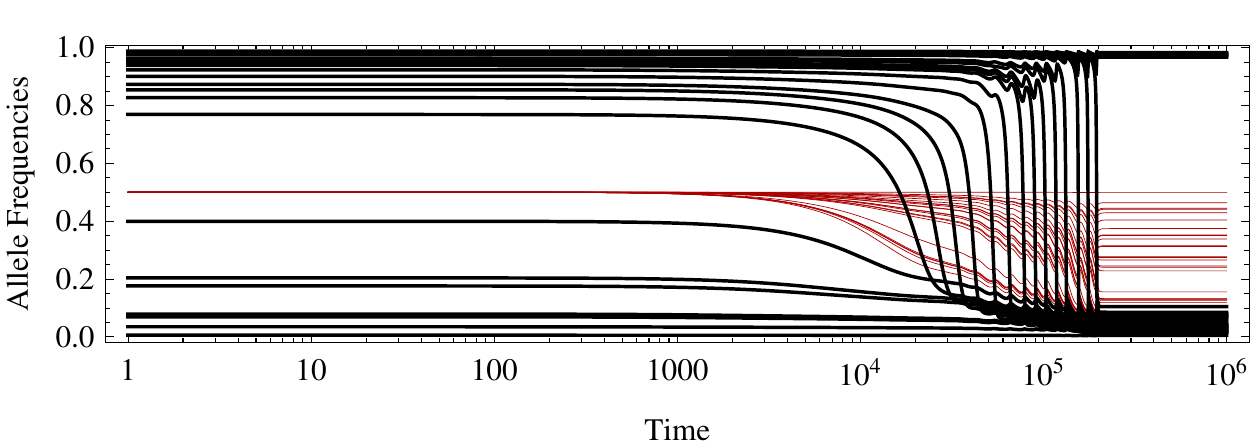}
\caption{Response to a gradually shifting optimum of a polygenic trait. (A) Deviation of the trait mean from the newly positioned optimum. (B)  Black: exact numerical results. Dashed black line: house of cards prediction,  ($\nu = 2 n\mu/S$). Dotted black line: exact value from Eq. \ref{Eq:GeneticVarianceMain}. (C) Response of the allele frequencies; black lines: alleles of large effect; thin red lines (color online): alleles of small effect. The optimum linearly  moves from $-z_\circ$ to $z_\circ$ between $t=0$ and $t=10^4$, and afterwards stays constant at $z_\circ$. Other parameters as in Fig. \ref{Fig:OptimumShiftZV}.}
\label{Fig:MovingOptimumZV}
\end{figure}

If the  optimum is shifted slowly enough, the deviation $\Delta z$ remains very small. We also see that the genetic variance then hardly
changes (for example, Fig. \ref{Fig:MovingOptimumZV}). After some time the population keeps evolving but reaches a stationary state.  If the optimum suddenly stops, the population settles at a state characterized by a lower genetic variance, but larger deviation from the optimum as in the case when it adapts to a rapid shift of the optimum, as in the previous section. How can we explain these patterns?

\begin{figure}
\begin{center}
\includegraphics[scale=0.7]{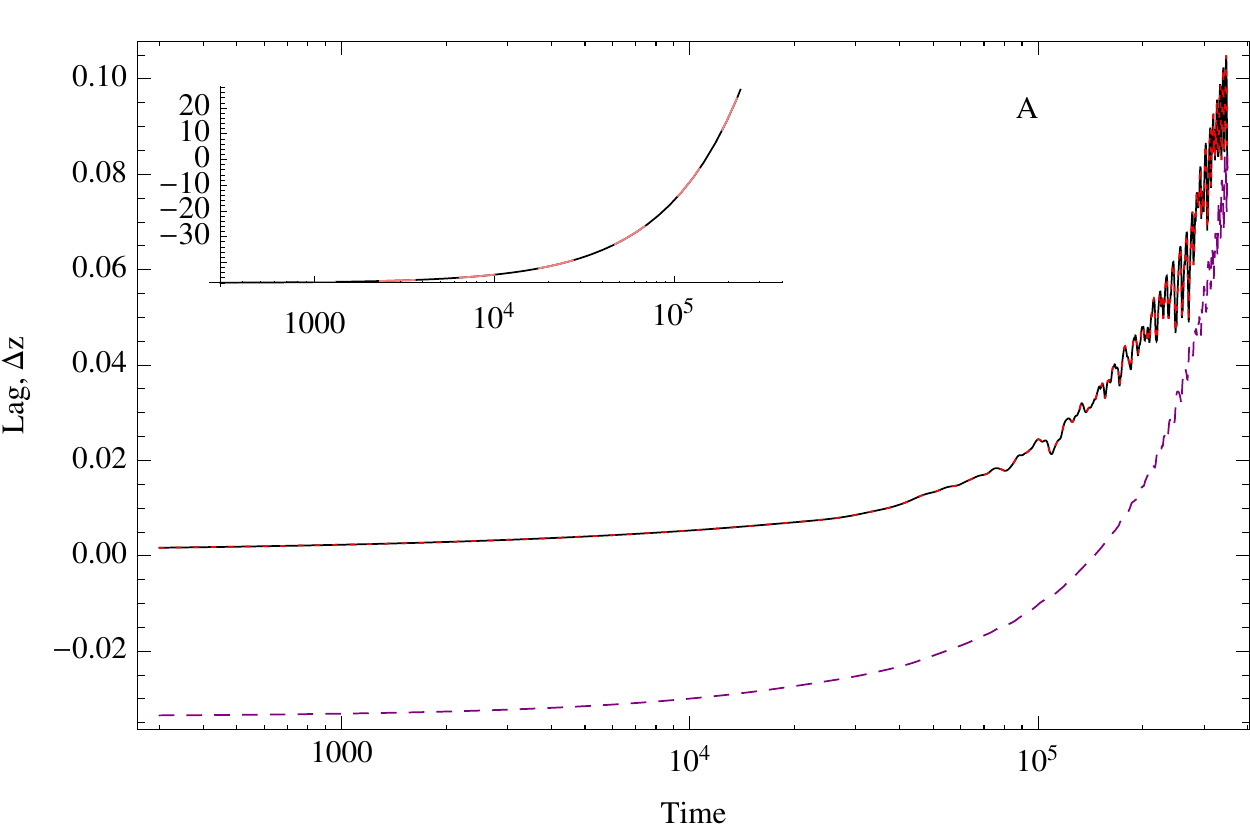}
\includegraphics[scale=0.7]{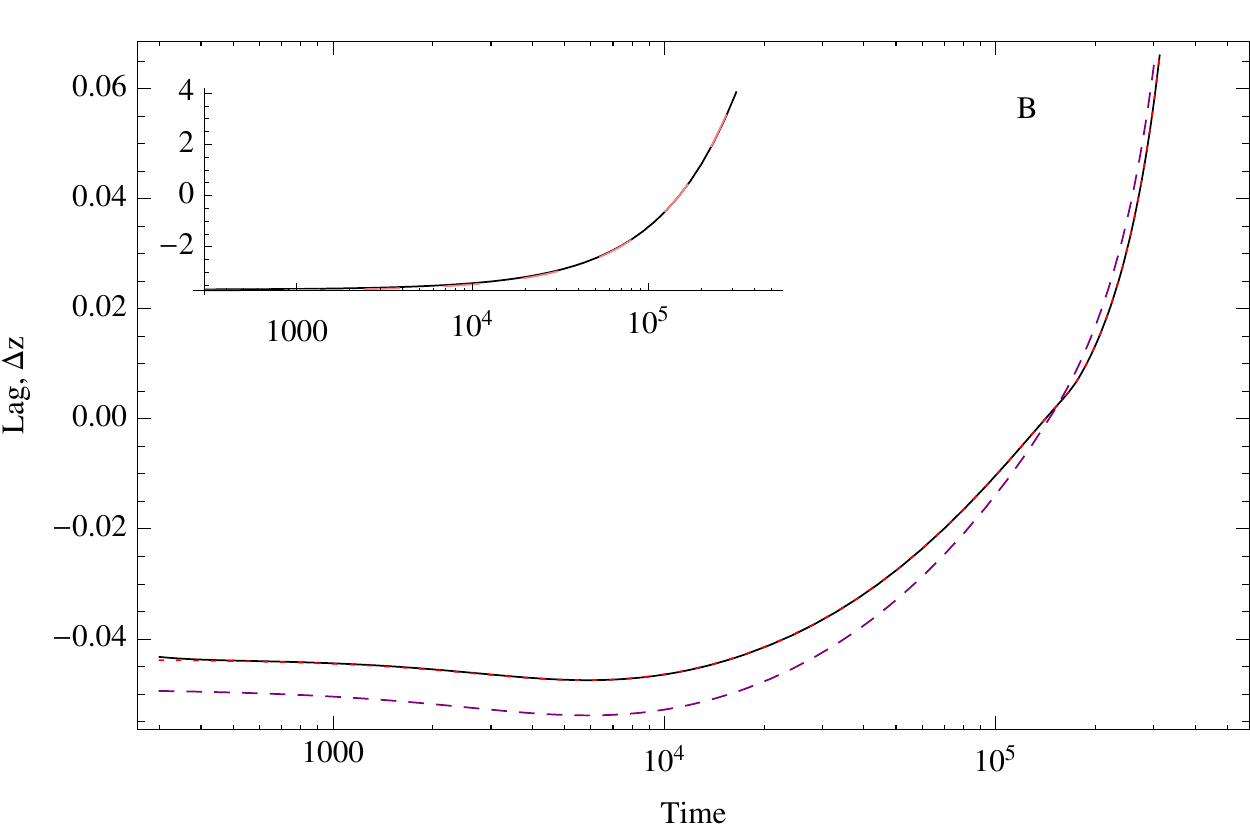}
\includegraphics[scale=0.7]{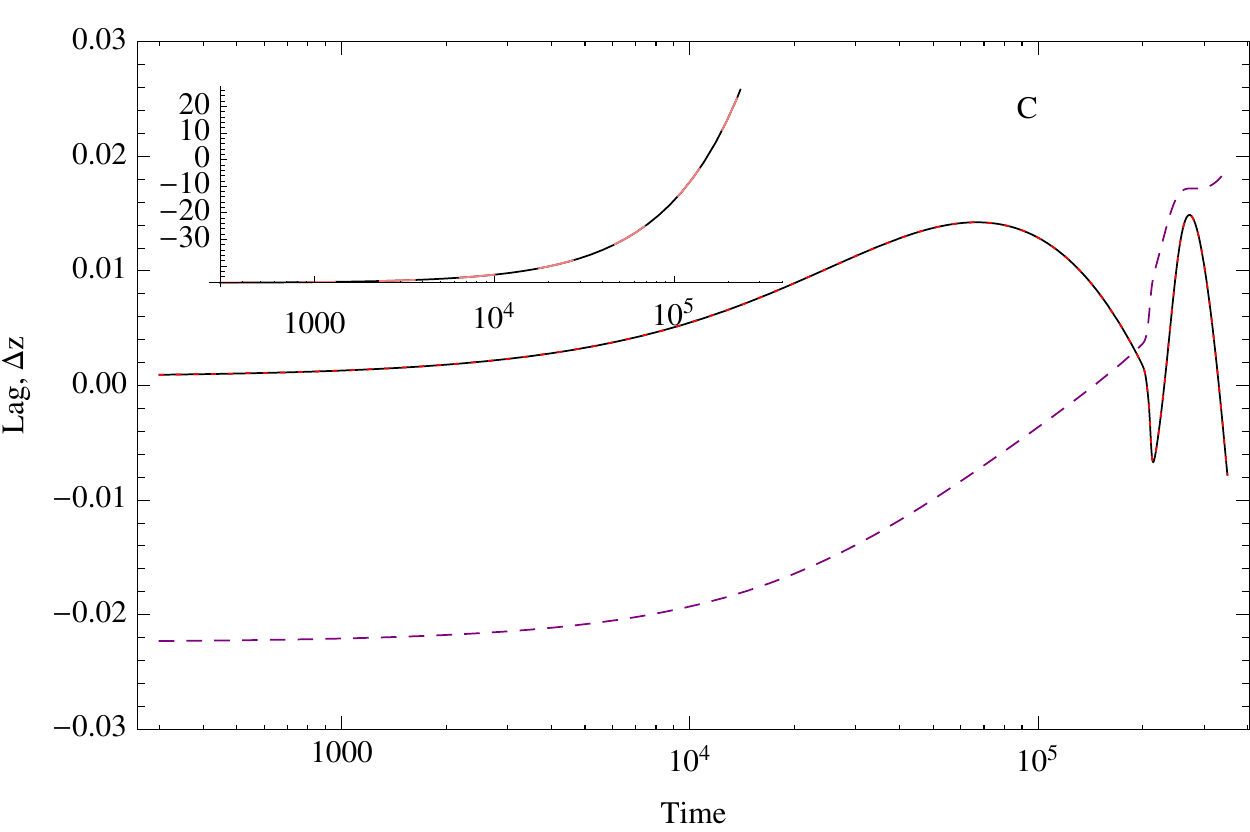}
\end{center}
\caption{Stationary lag approximation for the response of polygenic traits to a steadily moving optimum. In all cases the trait is is constituted by $n=1000$ loci. The optimum moves steadily shifting from -$\Omega$ to $\Omega$ in $T=300,000$ time units. The value of $\Omega$ was chosen to match the random initial condition. Black: lag $\bar{z}-z_\circ$; dotted red (color online) stationary lag $\Delta z^*$ (from Eq. \ref{Eq:Lag}); dashed purple (color online) stationary lag neglecting the third moment of the allelic effects. The insets compare $\bar{z}$ (black) and $z_\circ$  (dashed pink, color online) along all times. The trait is determined by (A) exponentially distributed effects (mean=0.1), $n_s = 435, \Omega=48$; (B) exponentially distributed effects (mean=0.01) $n_s=1000, \Omega=3.72$, and (C) Equal effects with mean $\simeq 0.1, n_s=0, \Omega=-46$. In all cases $S=0.1, \mu=10^{-4}$.}
\label{Fig:Lag}
\end{figure}

Under the infinitesimal model the population achieves a stationary lag from the optimum given by \( \Delta z^* = -\kappa /2S\nu\), where $\kappa$ is the speed of the moving optimum \cite{Lynch:1993vl,Jones:2004wk}. However, we see in Fig. \ref{Fig:Lag} that this approximation fails for finite number of loci of unequal effects.

Suppose that a moving optimum  changes linearly in time: $z_\circ(t) := \Omega_0 + ( \Omega_f- \Omega_0) t / T$. For simplicity we will consider optima starting at  $-\Omega$ and ending at $\Omega$. Hence the speed of the moving optimum is $\kappa=2\Delta\Omega/T$.

By summing Eq. \ref{Eq:AlleleFreqDynamics} over loci and using Eqns. \ref{Eq:TraitMean}-\ref{Eq:GeneticVariance}, we find that during transient evolution the deviation from the optimum is given by
\begin{equation}
\frac{d \Delta z}{dt} = -2 \nu \Delta z + S m_3 - 2 \mu \bar{z} -\kappa
\end{equation} 
where \(m_3 = \sum_i \gamma_i^3 p_i q_i (2p_i -1)\) is the third moment of the allelic effects. If the deviation from the optimum reaches a stationary state $\Delta z^*$ where  \( \frac{d \Delta z^*}{dt} = 0\), then:
\begin{equation}
\label{Eq:Lag}
\Delta z^* = \frac{\kappa + 2 \mu \bar{z} - S m_3}{2 S \nu}.
\end{equation}

Under the infinitesimal model, the breeding values are normally distributed which implies that \(m_3 =0 \). The genetic variance due to mutational effects is finite, but the mutation rate decreases with the inverse of $n$ and the term $\mu \bar{z}$ can be neglected. Consequently, Eq. \ref{Eq:Lag} reduces to the approximation of the infinitesimal model \cite{Lynch:1993vl,Jones:2004wk}. What limits are then necessary from the point of view of our model with a finite number of loci?

The stationary lag is not a constant; it represents a quasi-equilibrium state, and so we need to know how $\bar{z}, \nu$ and $m_3$ change in time. This is not feasible in an exact way, except under restrictive limits such as the infinitesimal model. Even under other simple assumptions, such as the HoC, predicting higher moments is hard \cite{Barton:1986vl,Barton:1987um,Burger:1991tr}. 

Figure \ref{Fig:Lag} shows that the third moment of allelic effects, $m_3$, is relevant for an accurate prediction of the lag. In fact, if all terms of Eq. \ref{Eq:Lag} are considered, there is virtually no distinction between the stationary lag approximation and the actual lag. However, neglecting the third moment does affect the prediction substantially. But the extent to which $m_3$ is relevant, depends on the distribution of effects.

Fig. \ref{Fig:Lag}B shows an example for a trait constituted only by alleles of small effects. The third moment is small, and neglecting it leads to a good approximation of the stationary lag. This is consistent with  the infinitesimal model as a limit of many loci of small effects. 

When traits are determined by alleles of large but equal effects the distribution of allelic effects is also asymmetric. As the optimum advances, traits with  unequal effects allow many small adjustments. These gradual changes allow fine tuning of the deviation from the optimum. Under unequal effects, this results in high frequency but low amplitude fluctuations of the lag. However, under equal effects the allele frequencies change in a coordinated fashion and the equilibria are more robust to deviations from the optimum. Thus, we observe fewer but larger fluctuations. 

Comparing Figs. \ref{Fig:Lag} A and B we find that traits with mixed allelic effects initially respond smoothly, but eventually enter a highly fluctuating phase. This does not happen when all the effects are small. We must point out that strong fluctuations can occur when the optimum is very close to the range of response of the trait. In the Supplementary Information 5 we show that different initial conditions converge to the same erratic trajectories, and show that these are not chaotic, but quasi-periodic (deterministic but unpredictable fluctuations of many different frequencies, characterized by a zero Lyapunov exponent).

The existence of the quasi-periodic phase explains why, if the optimum suddenly halts, the population remains stuck at a local optimum. This is because the populations are not able to wander freely in the fitness space. Being driven by the moving optimum, they are forced to stay in states that keep a certain deviation from it. In turn, in equilibrium, this poses some directional selective pressure, which biases even further the allele frequencies, resulting in a loss of genetic variation.
 
 In the Supplementary Information 5 we also study a moving optimum that oscillates smoothly with different frequencies and amplitudes. The lag enters a periodic phase of many frequencies (i.e. it is not smooth), and the fluctuations increase as the frequency and the amplitude increase.  We also study damped oscillations. Surprisingly, once the oscillations stop the population end up better adapted when compared with linearly moving optima.

\section*{Discussion}

Our analyses help to understand the relative contributions of alleles of large and of small effect in the maintenance of genetic variation and in the response to stabilizing selection. Interestingly, our analysis questions whether the specific distribution of allelic effects is relevant.  Naturally, a more robust interpretation of these results requires understanding how genetic drift affects the  distribution of allele frequencies. In this model, alleles of small effect are at intermediate frequencies. However these will be fixed by genetic drift. This could induce major changes to our results when there are many alleles of small effect. Thus, selection on many traits and genetic drift might change the picture substantially by maintaining larger variation, even though the details of the distribution of allelic effects might not be relevant.

Under genetic drift, the eventual fate of any allele is fixation or loss. Although drift may seem an additional complication, it also has an interesting effect, namely to allow access to parts of the fitness landscape that were inaccessible from a given state of a deterministic population \citep{deVladar:2011p5333}. In this sense, genetic drift smooths the landscape, and although stochastic effects are introduced, the expected trajectories are somewhat regularized, because the populations can easily escape suboptimal peaks \citep{Wright:1935vj,Barton:1989uh}, converging to fitter states. In this sense, drift aids adaptation, allowing alleles to jump across peaks by mutation and genetic drift \citep{Wright:1931,Wright:1932p356,Coyne:1997va}.

The stochastic HoC \citep[P. 270]{Burger:2001wy} shows that, on average, each locus at equilibrium contributes to the genetic variance by \(2\gamma ^2N  \mu \left/\left(1+\gamma ^2N  S \right)\right.\). Thus, the expected genetic variation, \(\langle \nu \rangle\), is smaller than the \(\nu _{{HoC}}\). Alleles of small effect will be fixed by drift, but their contribution is still small compared to alleles of large effect. Consequently, on average, the response of the trait is slower in finite populations than in infinite populations, which was already observed by us for the case of equal effects  \citep{deVladar:2011p5333}. However, in a strong selection regime, i.e. \(\gamma^2N  S>>1\), even alleles of small effect ($gamma^2N S<\mu$) will be near fixation. Thus for sufficiently large populations, the HoC approximation should hold. However, the problem is far from trivial because these fixed alleles of small effect can further induce deviations of the trait mean from the optimum.

These analyses assume that at equilibrium the trait is well adapted. Other factors such as asymmetric mutation rates can maintain a deviation from the optimum \citep{Charlesworth:2013jx}. In this case, a stationary population effectively experiences directional selection, and maintains even more genetic variance than when the trait matches the optimum. In our analyses we find in some cases that when traits deviate from the optimum, there is more genetic variance. Under asymmetric mutation rates, as in Charlesworth's model, the deviation from the optimum is maintained by two opposing forces: an asymmetric flux of mutations, and directional selection towards the optimum. In our model the populations simply stand at a suboptimal peak.

\subsection*{Selection on many traits and pleiotropy.} 
 It is often argued that stabilizing selection acts on multiple traits. Under a common polygenic basis, two traits that are subject to antagonistic selective pressures remain at an intermediate value that is a compromise amongst the optimal solutions. Alleles of large effect that are not subject to these pleiotropic effects can contribute significantly for the response to selection, even though when their contribution to the genetic variance is negligible due to the large number of loci \citep[ provide many examples]{Kelly:2009uh}. In this section, we show that our previous result are relevant in this larger context.
 
A more general calculation for many traits under selection shows that if several traits are all adapted to their optima, there are still two classes of alleles: near fixation and at intermediate frequency, but the criterion for locus $i$ to be near fixation is \(\Gamma_i \equiv \sum_k \gamma_{k i}^2 S_k > 4\mu\), where \(S _k\) is the selection strength on trait  $k$ and \(\gamma_{ki}\) is the allelic effect of locus $i$ on trait  $k$ (Appendix D). However, if the optimum for one trait favors alleles at the `+' state, and the optimum of the other trait favor alleles at the `-' state, then the net effect of selection on an allele might partly neutralize. In this case, deviations from the optima will exist and both alleles will be maintained at intermediate frequency, and the genetic variance in the population will be high, as they will contribute by $\gamma_{ki}^2/2$ even if  $\gamma_{ki} > \hat{\gamma}$.

For multiple traits that share a polygenic basis, only a few principal components will experience strong stabilizing selection; all the other components will be subject to only weak selection, otherwise the genetic load would be prohibitively large \citep{Barton:1990uo}.  Hence, it remains unclear when (and unlikely that) a particular focal trait is the main component of fitness \citep{Barton:1990uo}. 
Therefore, if the stabilizing nature of selection is attributed to pleotropic factors, the equilibrium genetic variance will be decreased  \citep{Barton:1990uo,Turelli:1985tl,Slatkin:1990wz}: if the strength of selection on the $M$ traits are the same we get that $\nu=\nu_{HoC}/M$. 

The observed differences in fitness can be due to other correlated traits, as explained above, or to pleiotropic effects directly affecting fitness. For morphological traits, the distribution of allelic effects is positively correlated with fitness effect  \citep{Keightley:1990vb}. However it remains difficult to disentangle whether pleiotropy or multivariate selection is the acting mode of fitness reduction \citep{Barton:1990uo,Zhang:2003ti}

Under antagonistic selection the picture is different. The HoC model for many traits \cite{Turelli:1985tl} shows that the genetic variance of one trait depends on the strength of selection of the other traits (even if the traits are uncorrelated). In this case each locus near fixation contributes by $2 \gamma_{1i} \gamma_{2i} \mu/\Gamma_i$. However, we must consider that the condition $\Gamma_i >4\mu$ is dependent not only on the distribution of allelic effects, but also on the distribution of selective coefficients $S$. If the latter has a mean of zero and small variance (weak selection), the fixation condition would be hard to fulfill, and most alleles will be at intermediate frequency leading to high genetic variance (consistent with \citet{Zhang:2003ti}).

\subsection*{Admixed populations and genetic incompatibilities.}
Suppose that two populations that are genetically differentiated come into contact. Will a subsequent admixture result on maladapted offspring? In SI 6 we show that the admixed population necessarily has larger genetic variance than the source populations, even if the latter have the same trait mean and variance. This is because the populations might be at different adaptive peaks that have the same or very similar phenotypic distribution. However, there are $\tilde{n}$ loci with distinct alleles in each population, which cause the excess variance relative to the parental mean. This will be caused by alleles of large effects, each one contributing by \( \gamma _i^2-2\tilde{n} \frac{\mu }{S} \). This can be interpreted as the expression of genetic incompatibilities between the two divergent populations, and emphasizes the role of stabilizing selection and epistasis in the process of speciation \citep{Barton:1989uh,Barton:2001ve}. (However, this mechanism is of a different nature than the paradigm of Dobzhansky-M\"uller incompatibilities). 
After secondary contact the population might develop isolation and re-adapt to its original state, retaining the incompatible alleles, or it might
hybridize and re-adapt to a new state. This will depend on the initial degree of admixture, but also on the otherwise negligible deviations from the optimum as well as on genetic drift, factors that we have not considered.

\subsection*{SNPs as genomic signatures of stabilizing selection.}

Under the assumptions of our analyses, most loci with high heterozygosity will have small effects, whereas alleles of large effect will have much lower heterozygosity, a result consistent with early results of the neutral theory \citep{Kimura:1969wt}. In turn, our results support the well-known idea that there can be substantial measurement bias in the estimation of allelic effects from QTL or GWAS: alleles of large effect will be harder to detect than polymorphisms of alleles with small effect. For instance, most effects that we can map are expected to be small. This is consistent with the  knowledge that most alleles have small effects. Furthermore, if alleles of large effect are common, our results indicate that they will be close to fixation, and thus rare in the population, and consequently less likely to be detected.

Ignoring drift and equating \(n_s\) to the SNPs on a genome of size $n$, and the proportion of fixed alleles to \(P=n_f/n\), and assuming that this proportion is homogeneous not only across the genome, but also across the set of loci that affect different traits (questionable suppositions of course), this implies that traits are approximately at a fraction $P$ of the total genetic variation, as we saw above. The ratio $n_s/n$ is on the order of 1:100 or 1:1000, thus  $P\simeq .99$. The evolution of these traits are mutation-limited rather than by standing genetic variability.  These estimations assume linkage equilibrium, as are the SNPs identified by GWAS, which are often spread across the genome. (Clearly, this does not apply within genes, coding or regulatory sequences, as linkage is tight). 

Different populations that show similar trait distribution and genetic variation may still differ at individual SNPs, especially if these have  large effects. Thus, a particular allele might not be uniquely associated with a particular trait, even if they are causally related.  This justifies and is consistent with GWAS findings that several variants can be associated with different alleles. What we have shown is that these causal variants are expected to contribute equally to the genetic variance, irrespective of the specific genetic makeup of the quantitative trait(s).

\subsection*{Acknowledgements.} The authors would like to thank Tiago Paixao and Daniel Weissman for the discussions. This project was funded by the ERC-2009-AdG Grant for project 250152 SELECTIONINFORMATION.



%
\section*{Appendix A: Bifurcation points}

The bifurcation points occur when there is a change in the stability of the equilibrium allele frequencies by varying a parameter. This means that, in addition to the equilibrium condition $dp/dt=0$, we also require that the eigenvalue vanishes at the point of equilibrium. If we rescale the equations in terms of $\delta$ and $m$, we have to solve the following system:

\begin{equation}
\begin{cases}
 m (1-2 p)-(1-p) p (2 \delta -2 p+1)=0 &  \\
 -2 \delta -2 m+2 p
   (2 \delta -3 p+3)-1=0 & 
\end{cases}
\end{equation}

By eliminating $p$ from the two equations we get that
\begin{equation}
8 \delta ^2 \left(2 \delta^2+2 (m-5) m-1\right)=(4 m-1)^3
 \end{equation}
 This formula defines the boundary in the diagram of Fig. \ref{Fig:PhaseDiagram} in the main text. Clearly. if there are no deviations from the trait, $\delta=0$, the right hand side gives the critical value $\hat{m} =1/4$. Also, as $m$ vanishes $\delta\rightarrow1/2$. In general, the last equation gives the boundary  for any arbitrary deviation.

We can also eliminate $m$ and get the allele frequency $p$ at which the bifurcation occurs, given by the solutions to the cubic:
\begin{equation}
 -8 p^3+4 (\delta +3)
   p^2+(-4 \delta -6) p+1+2 \delta=0
\end{equation}

Notice that in order to keep allele frequencies $0\leq p \leq 1$, it must be fulfilled that $-1/2 \leq \delta \leq 1/2$.

\section*{Appendix B: Perturbation analysis for small deviations from the optimum}
Consider a approximate solution to Eq. \ref{Eq:AlleleFreqDynamics} expressed as $p=P_0 + (\Delta z)  P_1 +  (\Delta z)^2 P_2 + O[(\Delta z)^3]$. The time derivative of $p$ neglecting terms of order $(\Delta z )^3$ is
\begin{equation}
\begin{array}{lcl}
\dot{p}&\simeq&\dot{P}_0+ (\Delta z)  \dot{P}_1  +  (\Delta z)^2  \dot{P}_2  \\
&=& (1-2P_0)\left[\mu -S\gamma^2 P_0 (1-P_0) \right] +\\
&&+(\Delta z) \left[ -S\gamma \left(2P_0 (1-P_0) (1-\gamma P_1)+\gamma P_1 (1-2P_0 )^2 \right) -2\mu P_1\right] + \\
&&+(\Delta z)^2 \left[ S \gamma  \left(\left(2 P_0-1\right) P_1 \left(3
   P_1+2\right)+6 \left(P_0-1\right) P_0
   P_2+P_2\right) -2 \mu  P_2\right]~.
\end{array}
\end{equation}

The unperturbed solutions for $P_0$ are those given in the main text, i.e. Eq. \ref{Eq:NoDevSolutions}. In equilibrium, we require the terms proportional to $\Delta z^m$ in the last equation to vanish, which gives the following two solutions for alleles of small and of large effects respectively:
\begin{equation}
P_1 = \left\{ 
\begin{array}{lc}
\frac{S \gamma}{S\gamma^2-4\mu} & \left( \gamma^2 < 4\mu/S \right) \\
\frac{2\mu/\gamma}{4\mu-S\gamma^2}  & \left( \gamma^2 > 4\mu/S \right)
\end{array} \right.
\end{equation}
Notice that both quantities are negative.

The second order perturbations give $P_2=0$ for alleles of small effect, and for alleles of large effect
\begin{equation}
P_2 = \pm \frac{4 \mu  \sqrt{\gamma  S} (\mu +\gamma 
   S)}{(4 \mu +\gamma  S)^{5/2}}
\end{equation}
where the sign indicates whether the allele is in the `+' or `-' state.

\subsubsection*{Deviations from the optimum.}
We can estimate the deviation from the optimum by summing over all alleles. This leads to a quadratic equation for $\Delta z$ with the following solution
\begin{equation}
\Delta z = \frac{1 -\zeta_1}{2 \zeta_2} \pm \left[ \left( \frac{1 -\zeta_1}{2 \zeta_2}\right)^2 -\frac{\zeta_0 - z_\circ}{ \zeta_2}  \right]^{1/2} ~,
\end{equation}
where the factors $\zeta_k$ are
\begin{equation}
\zeta_k = 2 \sum_{i=1}^{n} \gamma_i {P_k}_i 
\end{equation}
(${P_k}_i$ is the $k$'th perturbation at the locus $i$).

\subsubsection*{Maximum trait deviation.}
Taking the solution for alleles of large effect, we can calculate what is the maximum allowed deviation from the trait, $\tilde{\Delta} z$. For this, we equate the allele frequency to zero (for positive deviations) or equivalently to one (for negative deviations). Assuming positive deviations:
\begin{equation}
\min{p} = 0 = \frac{1}{2}\left[ 1- \sqrt{1-\frac{4\mu}{s\gamma^2}}\right] + (\tilde{\Delta} z) \frac{2\mu/\gamma}{4\mu-S\gamma^2}
\end{equation}
Which gives a deviation of
\begin{equation}
\tilde{\Delta} z = \gamma \left(\frac{S\gamma^2}{4\mu}-1 \right)\left[ 1- \sqrt{1-\frac{4\mu}{S\gamma^2}}\right]
\end{equation}
Notice that the larger effects tolerate larger deviations. Hence we need to take the minimum of all these deviations, to ensure stability for all alleles. This is thus given by the smallest allele of large effect. Assuming that  $S\gamma^2>>4\mu$  the expression above simplifies to
\begin{equation}
\tilde{\Delta} z = \frac{\gamma}{2}\left( 1 - \frac{3\mu}{S\gamma^2}\right) ~,
\end{equation}
and for alleles of extremely large effect, the deviations can be at most of order $\tilde{\Delta} z \simeq \frac{\gamma}{2}$, as reported in the main text.
\section*{Appendix C: Probability of allelic states}
In this appendix we derive the probability $\rho$  of finding an allele  at the `+' state, that is, Eq. \ref{Eq:DistAllelicState} on the main text.

As mentioned in the main text, we assume that the trait mean has a value $Z=\bar{z}=z_\circ$, and calculate the probability $\rho$ that given allele $X$ is at the `+' state. That is \(\rho \equiv \Pr [X=1 | Z = \bar{z} ] \). We first decompose this probability using Bayes' theorem: \[ \Pr [X=1 | Z = \bar{z} ] =\Pr[X=1] \frac{\Pr [ Z = \bar{z} | X=1] }{\sum_y \Pr [ Z = \bar{z} | X=y] Pr[X=y] } .\] 
Then, express the trait mean as the sum over loci $\bar{z}=\sum_i (2x_i-1)\sqrt{\gamma_i^2-4\mu/S} $, where $x_i$ indicates whether the allele is close to $x=1$ or $x=0$. Here we assumed that the background alleles are near fixation. Summarizing:
\begin{equation}
\begin{array}{ccl}
\Pr [ \bar{z} | x_j=1] &=& \Pr[ \sum_i (2x_i-1)\sqrt{\gamma_i^2-4\mu/S}  | x_j=1] \\
&=& \Pr [ \bar{z} - \sqrt{\gamma_j^2-4\mu/S}  ]
\end{array}
\end{equation}

Since the trait is a sum over independent loci, we can approximation that the trait distribution is  normal (central limit theorem). Its variance $V$, is given by summing over the background loci of large effect:
\begin{equation}
V =  \sum_{i\neq j} \left( \gamma_i^2- \frac{4\mu}{S}\right) ~.
\end{equation}
Now assume that the initial configurations between `+' and `-' alleles are chosen uniformly, i.e. $\Pr[X=1] = \Pr[X=0] = 1/2$. The sum in the denominator of Bayes' theorem involves only two gaussian terms. Putting the pieces together this leads to Eq. \ref{Eq:DistAllelicState}:
\begin{equation}
\rho_j =1\left/\left(1+ \exp\left[-2\frac{z_\circ}{V} \sqrt{\gamma_j^2-4\frac{\mu}{S} }\right] \right) \right. ~.
\end{equation}

To accommodate deviations from the optimum trait value, we proceed in a similar way, but using the first order perturbation on the allele frequencies. Notably, the variance V does not change, since all alleles are displaced proportionally to $\Delta z$. We arrive at the expression
\begin{equation}
\label{Eq:DistAllelicStateWithDeviations}
\rho_j =1\left/\left(1+ \exp\left[-2\frac{1}{V} \sqrt{\gamma_j^2-4\frac{\mu}{S} }\left(z_\circ +\frac{\Delta z S \gamma^2_j}{S\gamma^2_j -4\mu}\right)\right] \right) \right. ~.
\end{equation}

Notice that the term proportional to $\Delta z$ denotes the strength of the directional selection component on allele $j$. For alleles of very large effect ($S \gamma^2_j>>4\mu$, and the term is approximately $\Delta z$, which as we saw before is very small. Thus the deviations from the optimum affect mainly alleles of large effect that are closer to the critical value $\hat{\gamma}$.
\section*{Appendix D: Stabilizing selection on multiple traits}
We consider a simple extension of our model, where stabilizing selection acts independently on many traits. Call $\pmb{z}=(z_1,\ldots,z_m)$, an array of $m$ traits that are under selection $\pmb{S}=(S_1,\ldots,S_m)$, and  $\pmb{z}_\circ$ their corresponding optima. Calling $\mathcal{S}=\text{diag}(S)$, we define fitness as \(W_{\pmb{z}} = \exp\left[ - (\pmb{z}-\pmb{z}_\circ) \cdot \mathcal{S} \cdot (\pmb{z}-\pmb{z}_\circ)^T / 2\right)\), where ``$\cdot$'' represents inner product. In principle we could accommodate correlation selection in the model by allowing the matrix $\mathcal{S}$ to have non-zero off-diagonal elements, but we leave out that possibility at the moments. Under weak selection, mean fitness becomes
\begin{equation}
\bar{W}\simeq \exp\left[ -\frac{1}{2} (  \Delta \pmb{z} \cdot \mathcal{S} \cdot  \Delta \pmb{z}^T + \pmb{S}\cdot \pmb{\nu} ) \right]
\end{equation}
where \( \Delta \pmb{z} = (\bar{\pmb{z}}-\pmb{z}_\circ)$ is the vector of deviations from the optima and $\pmb{\nu}$ is the vector of genetic variances.

The trait means and genetic variances are
\begin{eqnarray}
\bar{z}_k = \sum_{i=1}^n \gamma_{k i}( 2p_i-1) \\
\nu_k = 2  \sum_{i=1}^n \gamma_{k i}^2 p_i q_i
\end{eqnarray}
where $\gamma_{k i}$ is the allelic effect of locus $i$ on trait $k$. We are assuming that all $n$ alleles contribute to $m$ traits. This can be relaxed by simply assuming that  $\gamma_{k i}=0$ for some $i,k$.

The equilibria for this system is given by
\begin{equation}
0=-p_iq_i\left[2 \beta_i -\Gamma_i (1-2p_i) \right]+\mu (1-2p_i)
\end{equation}
where
\begin{eqnarray}
\beta_i = \sum_k^m S_k \Delta z_k \gamma_{ki} \\
\Gamma_i = \sum_k^m S_k \gamma_{ki}^2
\end{eqnarray}
We find that the two quantities above take the role of the deviation from the optimum and allelic effects on the single trait model. In fact, Eq. \ref{Eq:CubicPolynomial} in the main text holds, when we define $\delta = \beta / \Gamma $ and $m = \mu/\Gamma$. Consequently, the critical points, the scaled equilibria, and their stability are the same.

\end{document}